\documentclass[conference]{IEEEtran}
\IEEEoverridecommandlockouts
\pdfoutput=1
\usepackage{cite}
\usepackage{amsmath,amssymb,amsfonts}
\usepackage{algorithmic}
\usepackage{graphicx}
\usepackage{textcomp}
\usepackage{xcolor}
\usepackage{tabularx}
\usepackage[draft]{minted}
\makeatletter
\let\@float@c@listing\@caption
\makeatother
\usepackage{graphicx}
\usepackage[hidelinks]{hyperref}

\input{minted-style.tex}

\newcommand{\chameleon}{ChameleonIDE}

\makeatletter 
\newcommand{\linebreakand}{%
  \end{@IEEEauthorhalign}
  \hfill\mbox{}\par
  \mbox{}\hfill\begin{@IEEEauthorhalign}
}
\makeatother 

\newcommand{\todo}[1]{}

\def\BibTeX{{\rm B\kern-.05em{\sc i\kern-.025em b}\kern-.08em
    T\kern-.1667em\lower.7ex\hbox{E}\kern-.125emX}}

\newcommand{\ignore}[1]{} 
    
\begin{document}

\title{\chameleon{}: Untangling Type Errors Through Interactive
  Visualization and Exploration
}

\author{\IEEEauthorblockN{Shuai Fu}
\IEEEauthorblockA{
\textit{Faculty of Information Technology} \\
\textit{Monash University}\\
Clayton, Australiay \\
\texttt{shuai.fu@monash.edu}}
\and
\IEEEauthorblockN{Tim Dwyer}
\IEEEauthorblockA{
\textit{Faculty of Information Technology} \\
\textit{Monash University}\\
Clayton, Australiay \\
\texttt{tim.dwyer@monash.edu}}
\and
\IEEEauthorblockN{Peter J. Stuckey}
\IEEEauthorblockA{
\textit{Faculty of Information Technology} \\
\textit{Monash University}\\
Clayton, Australia \\
\texttt{peter.stuckey@monash.edu}}
\and
\linebreakand 
\IEEEauthorblockN{Jackson Wain}
\IEEEauthorblockA{\textit{Faculty of Information Technology} \\
\textit{Monash University}\\
Clayton, Australia \\
\texttt{ORCID: 0000-0003-2006-3538}
}
\and
\IEEEauthorblockN{Jesse Linossier}
\IEEEauthorblockA{\textit{Faculty of Information Technology} \\
\textit{Monash University}\\
Clayton, Australia \\
\texttt{ORCID: 0000-0001-6782-7019}
}

}

\maketitle

\begin{abstract}


Dynamically typed programming languages are popular in education and the software industry. While presenting a low barrier to entry, they suffer from runtime type errors and longer-term problems in code quality and maintainability. Statically typed languages, while showing strength in these aspects, lack in learnability and ease of use. In particular, fixing type errors poses challenges to both novice users and experts. Further, compiler type error messages are presented in a static way that is biased toward the first occurrence of the error in the program code. To help users resolve such type errors we introduce \chameleon{}, a type debugging tool that presents type errors to the user in an unbiased way, allowing them to explore the full context of where the errors could occur. Programmers can interactively verify the steps of reasoning against their intention. Through three studies involving actual programmers, we showed that \chameleon{} is more effective in fixing type errors than traditional text-based error messages. This difference is more significant in harder tasks. Further, programmers actively using \chameleon{}'s interactive features are shown to be more efficient in fixing type errors than passively reading the type error output.
\end{abstract}



Dynamically typed programming languages such as JavaScript and Python have risen in popularity in recent decades \cite{chatley_next_2019}. These languages present a low barrier of entry, especially to beginner programmers: they require no type declaration, variable types or object structures can be modified dynamically, and functions can deal with dynamic input using ad-hoc polymorphism and runtime reflection. However, studies show that dynamically typed languages negatively affect development productivity \cite{kleinschmager_static_2012}, code usability \cite{mayer_empirical_2012}, and code quality \cite{gao_type_2017, ray_large-scale_2017, meyerovich_empirical_2013}. They are often found to produce error-prone code \cite{chen_empirical_2020, wang_empirical_2015,xu_python_2016} and require strong programmer discipline to avoid pitfalls \cite{chen_empirical_2020}. For these reasons, many modern dynamically-typed languages have introduced static typing annotations as part of the core language features in recent years (e.g.\ \textit{TypeScript}~\cite{microsoft_javascript_nodate} and \textit{mypy}~\cite{mypy_mypy_nodate}).

Functional programming languages have long enjoyed rigorous type systems and expressive type-level features. Techniques such as type inference and algebraic types have been standard practice for decades in functional languages such as ML and Haskell, and more recently in multi-paradigm languages, such as Rust and TypeScript. Various type system advances were introduced in Haskell and ended up in mainstream languages years or even decades after, leading many to consider Haskell the ``type-system laboratory" \cite{hudak_history_2007}.  Type classes, an implementation of generic programming, were introduced to Haskell in 1988~\cite{hudak_history_2007}, and now can be found in most popular languages such as C\#~\cite{bill_wagner_constraints_2022}, Java~\cite{oracle_generic_2022}, and TypeScript~\cite{microsoft_documentation_2022}.

One crucial challenge of programming in statically-typed languages is that type errors can sometimes be difficult to resolve~\cite{tirronen_understanding_2015, hage_solved_2020}. In particular, they may point to locations that are not the root  causes of the type error, expose errors in cryptic language, or provide misleading fixing suggestions~\cite{wu_how_2017}.

This paper introduces \chameleon{}, an interactive type debugging tool for Haskell. It can visualize the relevant context of a type error: where it happens or could have happened and which parts of the code cause it. In addition, \chameleon{} allows programmers to interactively explore all the parts of code where multiple types can be inferred and to resolve ambiguity. The most noticeable features are the type compare tool (Section \ref{sub:type-compare}), the candidate expression card (Section \ref{sub:candidate-expression}), and the deduction step (Section \ref{sub:deduction-steps}). These features are integrated into a debugging environment and can be enabled or disabled separately based on the programmers' preferences and debugging needs. \chameleon{} is open-source and is available at ~\cite{anonymous_chameleon_2022}.  

This paper makes the following contributions:
\begin{itemize}
\item We provide the design and implementation of the \chameleon{} to visualize the relevant context of a type error and allow programmers to explore and verify the error locations in small chunks interactively.  
\item {
    We report the results of three experiments designed to evaluate \chameleon{}.}
\end{itemize}

Our experiments showed that programmers using \chameleon{} fix type errors faster than with traditional text-based error messages. This difference is more significant when solving harder tasks. Further, programmers who actively use \chameleon{} interactive features fix type errors faster than simply reading the type error output. Although \chameleon{} is designed to work with
the Haskell language, we plan to extend the underlying ideas to work with other strongly typed languages, such as Rust or TypeScript..

\section{Motivation}
The design requirements of \chameleon{} are motivated by limitations of traditional type errors, as documented in a number of studies (e.g.~\cite{yang_improved_2000, hage_solved_2020}), but which we illustrate here with a few motivating examples. 

\begin{figure}
    \centering
    \includegraphics[width=\linewidth,trim=0mm 35mm 0mm 0mm]{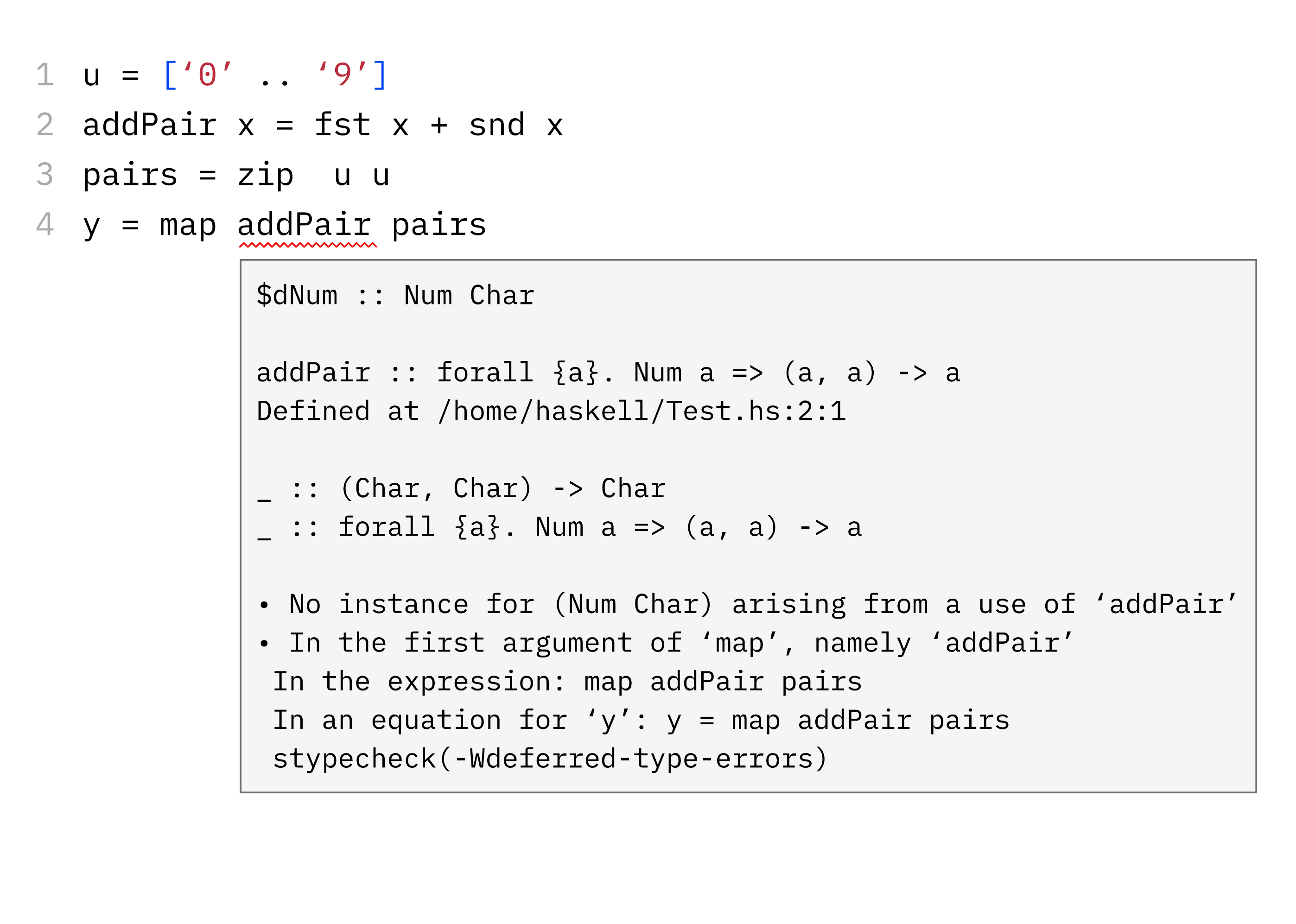}
    \caption{
    A type error displayed in Visual Studio Code\cite{microsoft_visual_nodate} and the Haskell Vscode extension\cite{haskell_haskell_nodate}.
The expression \texttt{addPair} is blamed for causing the type error. This may not match the programmers' intention. 
    }
    \label{fig:motivation-example}
\end{figure}
\subsubsection{\textbf{Traditional type errors show only limited location}}
Haack and Wells~\cite{haack_type_2004} noted that ``\textit{Identifying only one node or subtree of the program as the error location makes it difficult for programmers to understand type errors. To choose the correct place to fix a type error, the programmer must find all the other program points that participate in the error.}'' The type error in Fig.~\ref{fig:motivation-example} can be fixed in multiple locations. For instance  replacing \texttt{['0'..'9']} on line 1 with \texttt{[0..9]}, or replacing \texttt{fst x} and \texttt{snd x} on line 2 with \texttt {read (fst x)} and \texttt{read (snd  x)}. In the type error message, only the \texttt{addPair} expression on line 4 was blamed.  In this small example, the whole context is visible, but it can become problematic in large programs where the lines contributing to the type error are far apart in the source code.

\subsubsection{\textbf{Traditional type errors are biased}}
A common form of bias happens when a type error is reported in one expression, but it can occur in multiple other expressions as well. In Fig.~\ref{fig:motivation-example}, the error message arbitrarily focuses on only \texttt{addPair}, while ignoring that the literals in the definition of \texttt{u} may be incorrect. 
Another form of bias is that traditional type errors are often framed as conflicts between \texttt{Expected type} and \texttt{Actual type}. This framing is standard practice in most typed languages. However, what is \texttt{expected} and what is \texttt{actual} are a side effect of different unification orders rather than the intention of the programmer. In both forms, the error message may lead programmers to falsely believe the validity of parts of code and wrongly accuse others.

\subsubsection{\textbf{Traditional type errors give poor explanations}}
When the compiler rejects a program, the internal state of type checking is the result of a complex computation. But the details of this process are hard to explain to users and are usually not reported by compilers. For the typical type error shown in Fig.~\ref{fig:motivation-example}, the evidence for the type error is gathered from the previous declarations. These have to be rediscovered by programmers using less rigorous methods. 

\subsection{Design Goals of \chameleon{}}
Based on the limitations of traditional type errors, we give the following design requirements for \chameleon{}:

\noindent\textbf{Show} all the possible locations where the type error happened or could have happened.

\noindent\textbf{Explain} type errors avoiding jargon and internal constructs of the type checker.

\noindent\textbf{Do not presume} which expression is to blame for the type error based on the order of computation or which possible type for an expression is `actual' or `expected'.

\section{Chameleon IDE} \label{chameleon}
\begin{figure*}
    \centering
    \includegraphics[width=\textwidth, trim=0mm 10mm 0mm 0mm]{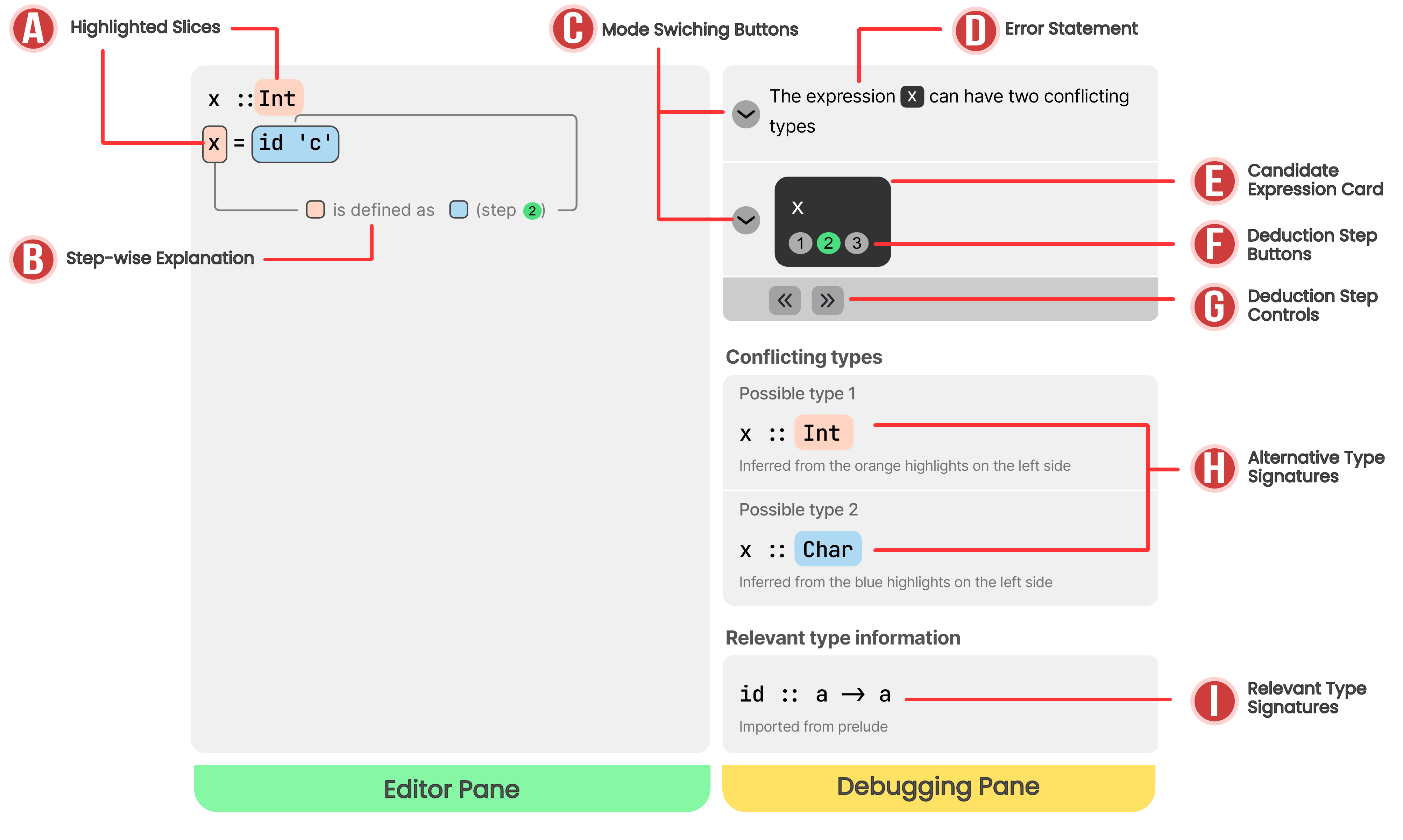}
    \caption{
        \textbf{The anatomy of \chameleon{}.}
        The editor pane (left) is similar to a traditional code editor. Fragments of source code may have a highlight
        color (A). Additionally, an explanation layer (B) displays if deduction steps are enabled. The debugging pane contains three blocks. First, the error statement block contains an error statement (D), optionally, a list of candidate expression cards (E), a list of deduction steps (F), and a control bar (G) to increment/decrement deduction step. Second, the conflicting types block shows two alternative types (H). Third, the relevant type information block shows additional information (I) that may help understand type errors.
    }
    \label{fig:anatomy}
\end{figure*}

\chameleon{} comprises two parts: a type inference engine and a novel interactive debugging interface. 
The debugging interface is designed from the ground up; the type inference engine is a re-implementation of the original Chameleon with several novel improvements, as described in Section \ref{sec:typeinferenceengine}.

\todo{I wonder about a high level overview diagram of tool / usage of tool, then the details of each component?}

\todo{Can you use this example in Motivating Example sectoon (see my comment above) then use it to explain tool in this section?  Make sense?

Is this example too simple...?}

\subsection{The Debugging Interface}

The \chameleon{} debugging interface provides three main features to visualize and explain type errors.


\paragraph{Type compare tool} \label{sub:type-compare}

\begin{figure}
    \centering
    \includegraphics[width=\linewidth, trim=0mm 10mm 0mm 0mm]{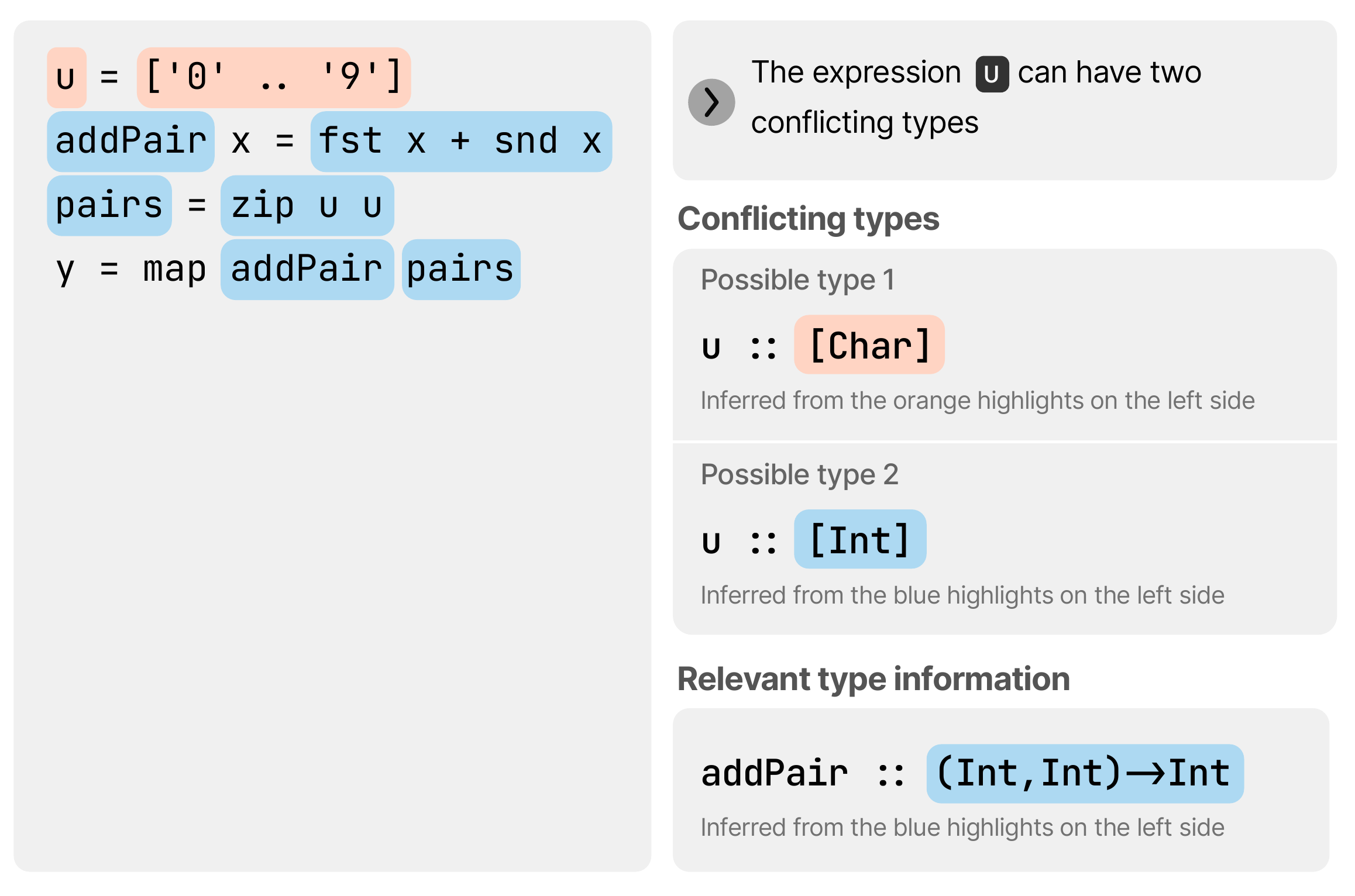}
    \caption{
        \textbf{\chameleon{} with type compare tool enabled}. \chameleon{} identified the conflicting types for the expression \texttt{u} and associated the relevant locations with each type. Compare the output with the traditional type error message in fig. \ref{fig:motivation-example}.
}
    \label{fig:compare}
\end{figure}

The type compare tool shows conflicting types in different colors, each type associated with one or more error locations highlighted in a matching color (Fig.~\ref{fig:compare}).  
If the programmers know the expression's intended type (they usually do), they will be able to eliminate half of the possible locations. 
A hover interaction over one of the possible types facilitates such bisection, causing only the relevant locations that contribute to that type to be highlighted. 

\begin{figure}
    \centering
    \includegraphics[width=\linewidth, trim=0mm 10mm 0mm 0mm]{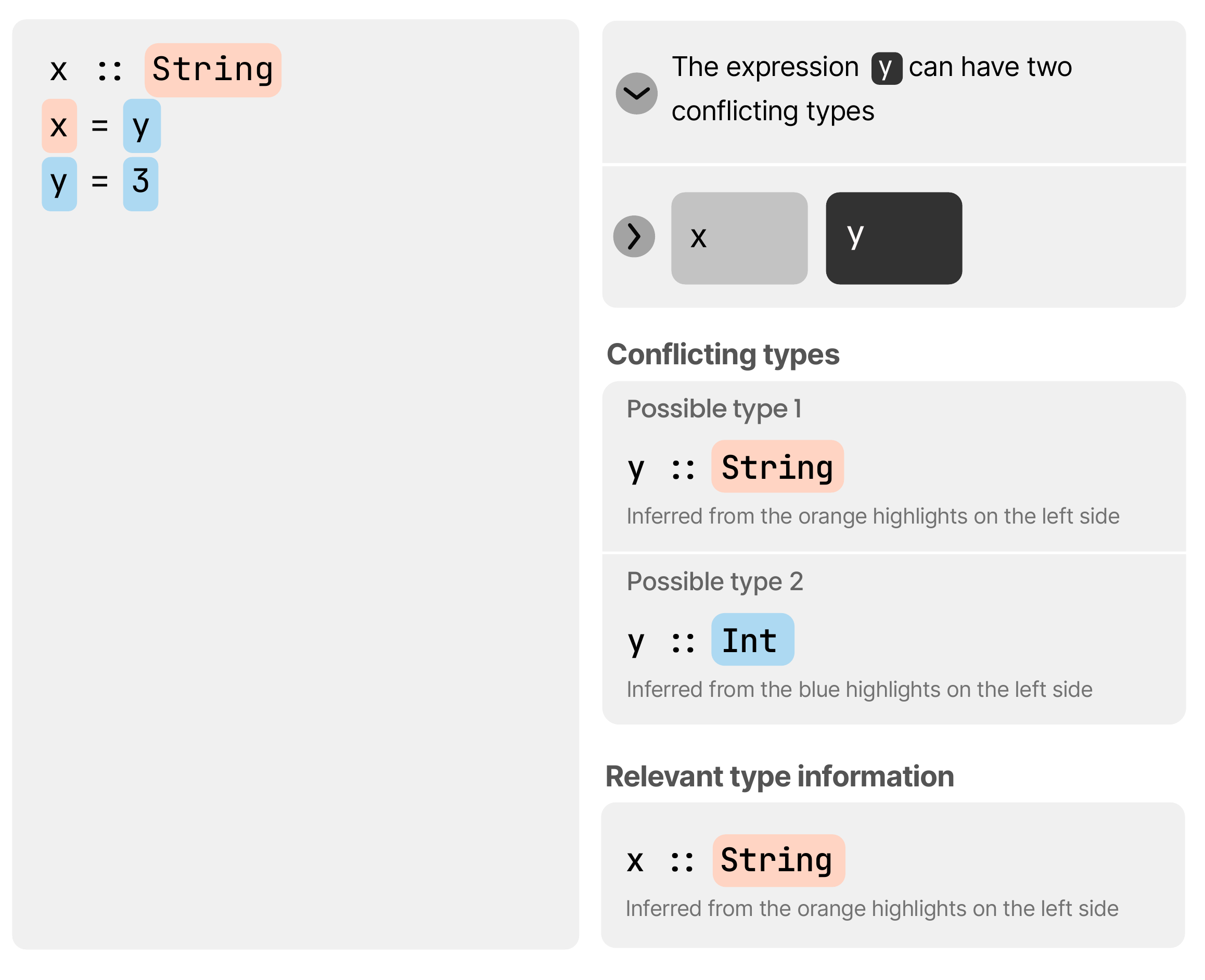}
    \caption{
        \textbf{\chameleon{} with candidate expression cards enabled.}
        Indicates the type error can occur in the definition of \texttt{x} or \texttt{y}.
    }
    \label{fig:expression}
\end{figure}

\begin{figure}
    \centering
    \includegraphics[width=\linewidth, trim=0mm 10mm 0mm 0mm]{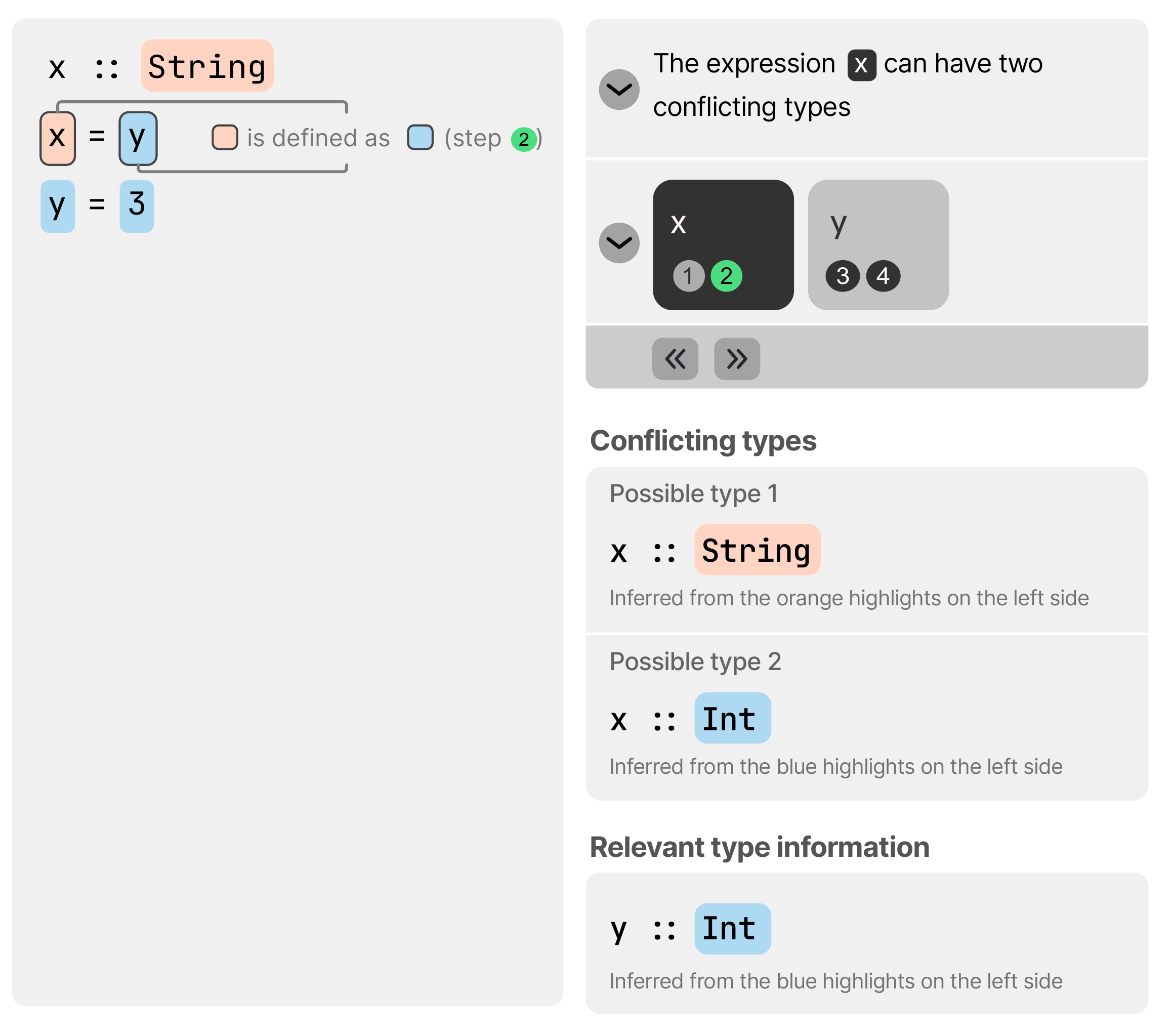}
    \caption{
        \textbf{\chameleon{} with deduction steps enabled.}
        \chameleon{} explains the type error in four steps. In the screenshot, the active step is step 2, where \chameleon{} shows that the expression \texttt{x} and \texttt{y} should have the same type. 
    }
    \label{fig:deduction}
\end{figure}

\paragraph{Candidate Expression Cards}  \label{sub:candidate-expression}

A candidate expression is an expression that can be inferred to have two conflicting types. 
When a type error is detected, \chameleon{} provides a list of all candidate expressions, and programmers are free to choose the problem to resolve by clicking on one candidate expression card. In the example shown in Fig. \ref{fig:expression}, \texttt{x} and \texttt{y} are both candidate expressions. Fixing either type error can make both expressions well-typed.


Programmers select a candidate expression card by clicking on one card. Once a card is selected, the information in the conflicting types block changes to reflect the change of candidate expression. In the editor pane, some error locations change highlight colors based on the updated candidate expression. Alternatively, programmers can preview the change of a candidate expression by hovering on one card. The hover effect is reverted once the cursor moves away.

\paragraph{Deduction steps}  \label{sub:deduction-steps}



\begin{figure}
    \centering
    \includegraphics[width=\linewidth, trim=0mm 0mm 0mm 0mm]{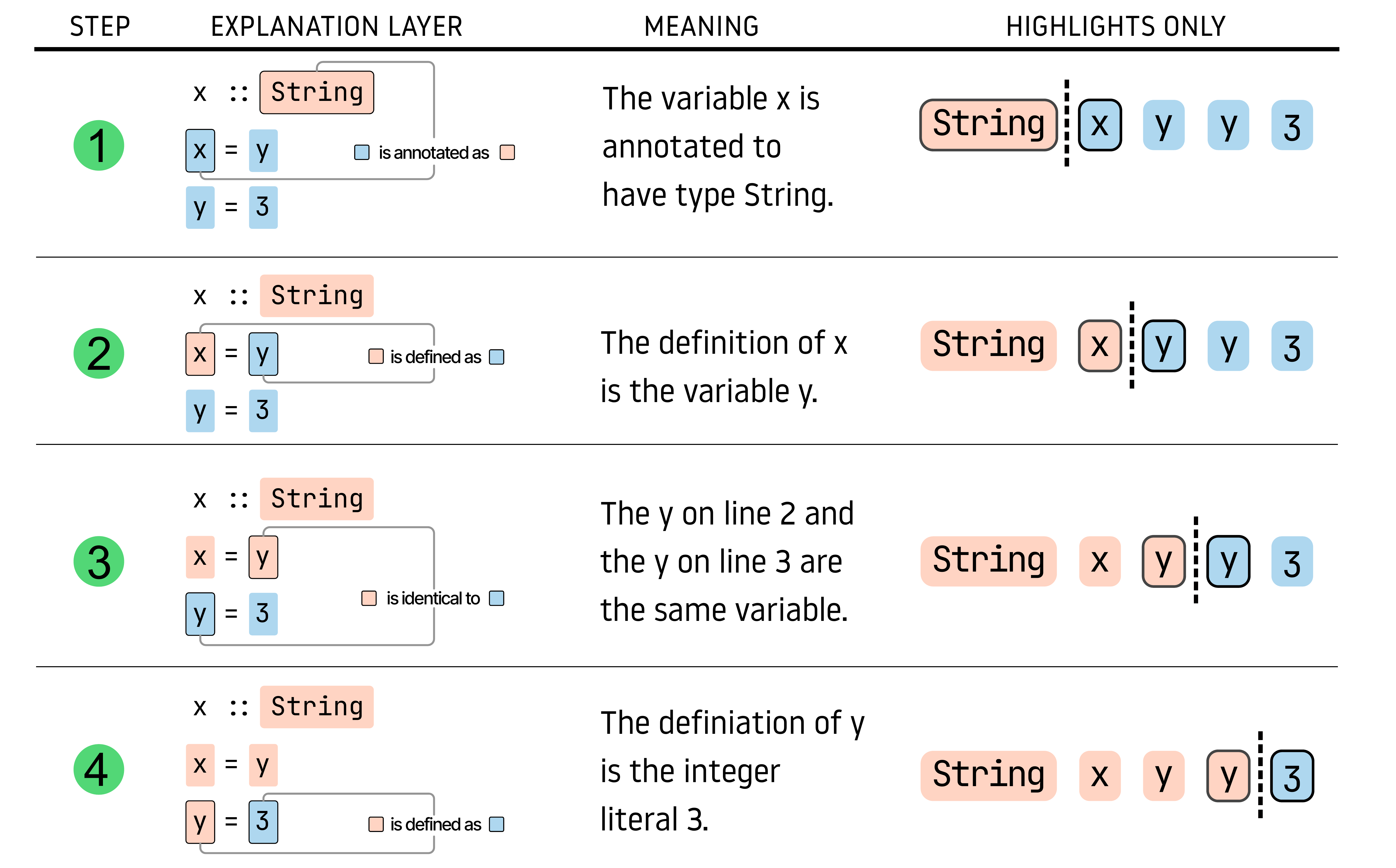}
    \caption{
Deduction steps if they are shown all at once. In practice, steps are shown one at a time. Programmers increment or decrement the step number using the step control bar (Fig. \ref{fig:anatomy}-G) or by directly clicking on a step button (Fig. \ref{fig:anatomy}-F). To increment or decrement the deduction step can be intuitively thought of as moving the position of the \textit{splitting point} (dotted lines) where the blue and orange highlights divide.
        }
    \label{fig:step-interface}
\end{figure}

Deduction steps allow programmers to explore all the error locations one at a time (Fig. \ref{fig:deduction}). Steps are shown as a list of sequentially  numbered circular buttons (step buttons) and an explanation layer in the editor window. In the explanation layer, the two locations under examination are outlined, and a line is drawn to connect these two locations. This line is accompanied by a human-readable text explanation of their semantic connection. Programmers are free to activate any step. The active step is shown in green. When activating a step, some highlights switch color. The message in the explanation layer changes accordingly. A program in Fig. \ref{fig:deduction} generates a list of steps shown in Fig. \ref{fig:step-interface} left.

Programmers can use mouse and keyboard shortcuts to increment or decrement the step number or jump to any step. Programmers resolve type errors by navigating through all the deduction steps and verifying whether each explanation aligns with their intention. Eventually, they will find a step that does not match, and the type error can be fixed by modifying one of the two outlined locations.

Internally, deduction steps are different ways to divide the error locations into two  groups, denoted by the two colors. Each color infers a different type of the candidate expression. Each increment/decrement of the step changes the splitting point (dotted lines in Fig.~\ref{fig:step-interface}) of the two colors.

\paragraph{Multiple Modes}

Nielson pointed out  that the two most important issues in designing for usability are understanding the users' tasks and the differences in users \cite{jakob_nielsen_usability_1993}. From analyzing how users use \chameleon{}, we realized that the ideal debugging interface should adapt to the specific programmer and programming task. There are cases where a programmer wants the debugger to simply ``show the answer", and others to dive deeper into the problem domain and search for the optimal solution. To accommodate the need to customize the level of information density and granularity of control, \chameleon{} provides three modes: basic, balanced, and advanced. Programmers can switch between modes by clicking on the mode switching toggles (Fig. \ref{fig:anatomy}-C). The features accessible from different modes are summarized in table~\ref{tab:chameleon-features}.

\begin{table}
    \centering
\begin{scriptsize}
\begin{center}

    \begin{tabular}{ l l  }
     \textit{Mode} & \textit{Features} \\ \hline
     Basic Mode & Type Compare Tool \\ \hline
     Balanced mode & Type Compare Tool \\
     & Candidate Expression Cards \\  \hline
     Advanced mode & Type Compare Tool \\
     & Candidate Expression Cards \\
     & Deduction Steps \\
    \end{tabular}
    \end{center}
\end{scriptsize}
    \caption{\chameleon{} modes and features}
    \label {tab:chameleon-features}
\end{table}

\subsection{The Type Inference Engine}
\label{sec:typeinferenceengine}

Chameleon was originally a command-line tool developed in the early 2000s to improve type error reporting 
for the Haskell programming language.
Unlike traditional type errors produced by the Glasgow Haskell Compiler (GHC)~\cite{ben_gamari_home_2022}, which uses a Hindley–Milner type inference system, Chameleon infers types using constraint solving. In Chameleon, constraints are generated from the source code based on typing rules. In addition, each constraint is labeled with the location where it is generated. This set of constraints is consistent if the program is well-typed and inconsistent otherwise. When a type error occurs, an efficient algorithm is used to derive a minimal subset of the constraints that still contain inconsistencies. This subset is called a Minimal Unsatisfiable Subset (MUS). From this, Chameleon can report a list of locations, using the labels of constraints that are in the MUS. Stuckey et al.~\cite{stuckey_interactive_2003} showed that program locations linked to the constraints from an MUS are all relevant to the type error and must include the cause of the error.

Despite successfully borrowing the underlying ideas, we could not reuse the original implementation of Chameleon since the project language standard and libraries used were out of date. 
Our \chameleon{} implementation extends the original Chameleon approach in a number of ways.

\paragraph{Recovering concrete types from type errors}

\begin{figure}
    \centering
    \includegraphics[width=\linewidth, trim=0mm 6mm 0mm 0mm]{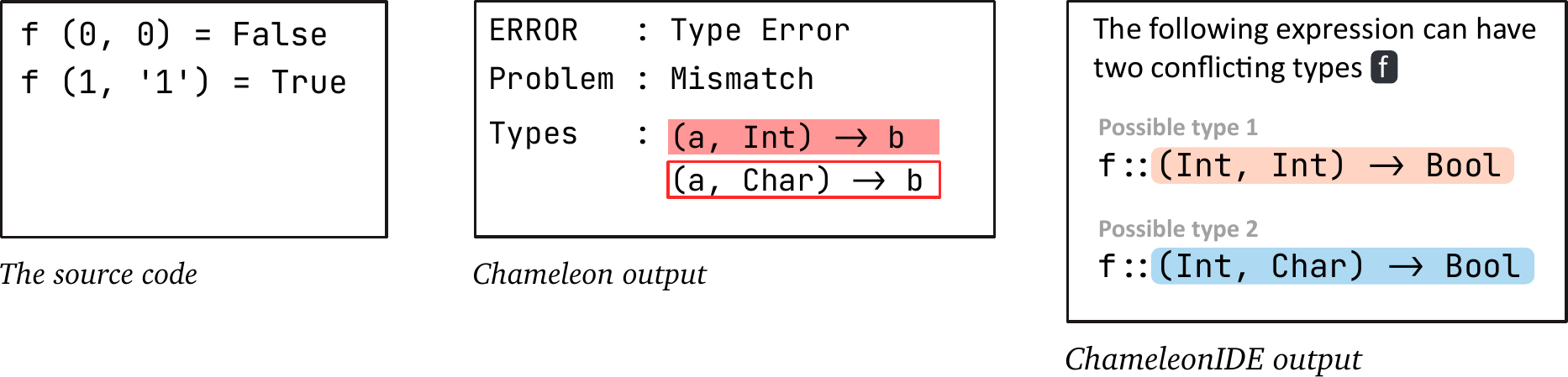}
    \caption{
Reporting the same type error, Chameleon uses more abstract types
\texttt{Int -> a} and \texttt{Char -> a}, while \chameleon{} uses the 
concrete types (types that do not contain type variables) \texttt{Int -> Bool} and \texttt{Char -> Bool}.
    }
    \label{fig:compare-to-original}
\end{figure}

Using only constraints from the MUS is sufficient to locate the type error, but to recover types from type errors we need constraints from parts of the program that are irrelevant to the type error.  For instance, consider an ill-typed 2-tuple where two possible types can be assigned: \texttt{(Int, Int)} and \texttt{(Int, String)}. The types reconstructed from Chameleon may be \texttt{(a, Int)} and \texttt{(a, String)}. Although the recovered types are theoretically correct, they introduce the notation \texttt{a}, which denotes a generic type variable that can be any type, making the error message harder to understand. To solve this issue in \chameleon{}, for each constraint \texttt{c} in the MUS, we find a maximally satisfiable subset (MSS) from all the constraints that contain every other element of MUS but not \texttt{c}. These maximally satisfiable subsets, while not helpful in error localization, will produce the most concrete types, see Fig.~\ref{fig:compare-to-original}. Concrete types, such as \texttt{Int} and \texttt{String},  often provides extra information to programmers. With a type of \texttt{(Float, Float)}, programmers may want to convey a point in 2d space. However, a type of \texttt{(a, Float)} does not preserve such information.

\paragraph{Type error explanation}

In addition, \chameleon{} provides support for type explanation. Similar to the type explanation system  in ~\cite{jun_explaining_2002},  \chameleon{} is able to produce a human-readable explanation, but for type errors. This is achieved by annotating nodes in the abstract syntax tree with constraints and the type inference rules used. We generate an inference history from constraints and accompanying annotations.

\definecolor{bg}{rgb}{0.95,0.95,0.95}
\begin{listing}[!ht]
\begin{Verbatim}[commandchars=\\\{\}]
    \PYG{k+kr}{if} \PYG{n}{a} \PYG{k+kr}{then} \PYG{n}{b} \PYG{k+kr}{else} \PYG{n}{c}
    \PYG{n}{a} \PYG{o+ow}{=} \PYG{l+s}{\PYGZdq{}True\PYGZdq{}}
\end{Verbatim}
    
\vspace*{-2mm}
\caption{A simple program that is ill-typed. It generates two constraints from line 1 and one constraint from line 2. }
\label{listing:1}
\end{listing}
For instance, for the program in Listing~\ref{listing:1}, \chameleon{}  generates the following constraints and labels (in brackets) $T_a = Bool$ (if condition), $T_b = T_c$ (if branches), $T_a= String$  (definition). Clearly, as $T_a$ can not unify with both \textit{Bool} and \textit{String}, this program is not well typed. \chameleon{} can construct a human-readable explanation from the MUS. An example output for Listing~\ref{listing:1} can be: \texttt{a} has type \texttt{Bool} because \texttt{a} is the condition of an if statement; however, \texttt{a} has type \texttt{String} because \texttt{a} is defined as the string literal \texttt{"True"}. This explanation facilitates the deduction steps (Section \ref{sub:deduction-steps}). \looseness=-1

\section{Walkthrough} \label{sec:walkthrough}
In this section, we showcase \chameleon{} by walking through examples of its
use. The examples are given from the perspective of a hypothetical Haskell 
programmer Maxine. 


\subsection{Basic mode} \label{sub:basic}
\begin{figure}
        \centering
        \includegraphics[width=\linewidth,trim=0mm 8mm 0mm 0mm]{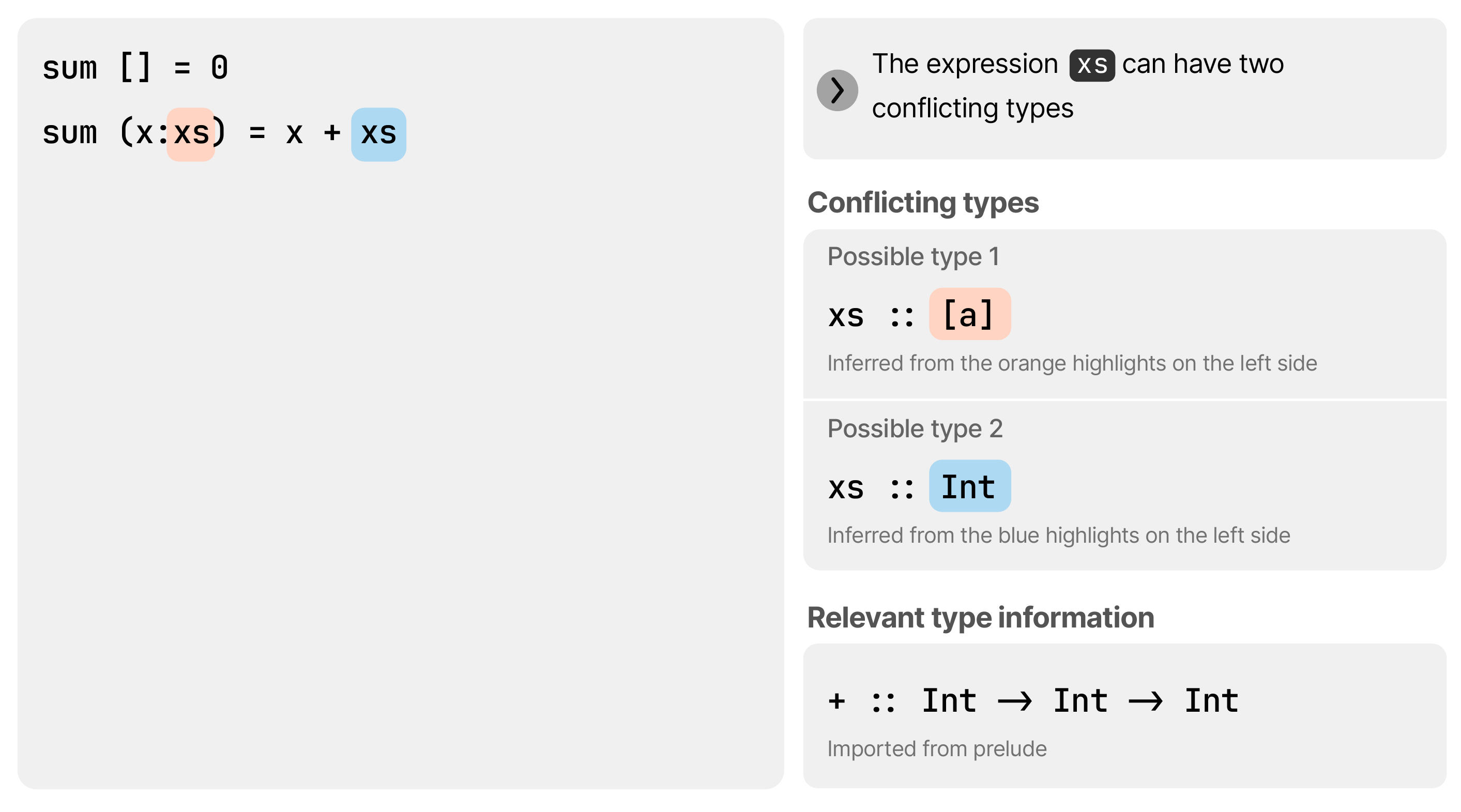}
        \caption{
            Maxine's code to calculate the sum of a list of integers;
            \chameleon{} reports an error on the expression \texttt{xs}.
            }
            \label{fig:basic-mode-1}
\end{figure}

Maxine writes a function to calculate the sum of a list of
numbers, but \chameleon{} shows there is a type error (Fig.~\ref{fig:basic-mode-1}). 
After reading the error reports, Maxine realizes that the error revolves 
around the expression \texttt{xs}. That is: \texttt{xs} can be
either \texttt{[a]} or \texttt{Int}. By matching the color in the
conflicting type block (Fig. \ref{fig:anatomy}-H) and the highlighted error locations 
Maxine knows that the \texttt{[a]} results from the pattern matching of the
\texttt{:} operator, while \texttt{Int} results from using \texttt{+} to
 add two expressions. 

At this point, Maxine knows the possible type 1 aligns with her intention, and therefore, the error locations with blue highlights must be erroneous. After examining the program, it comes clear that Maxine forgets to apply the \texttt{sum} function recursively at the right-hand side of the addition.

\subsection{Balanced mode} \label{sub:balanced}

\begin{figure}
        \centering
        \includegraphics[width=\linewidth,trim=0mm 8mm 0mm 0mm]{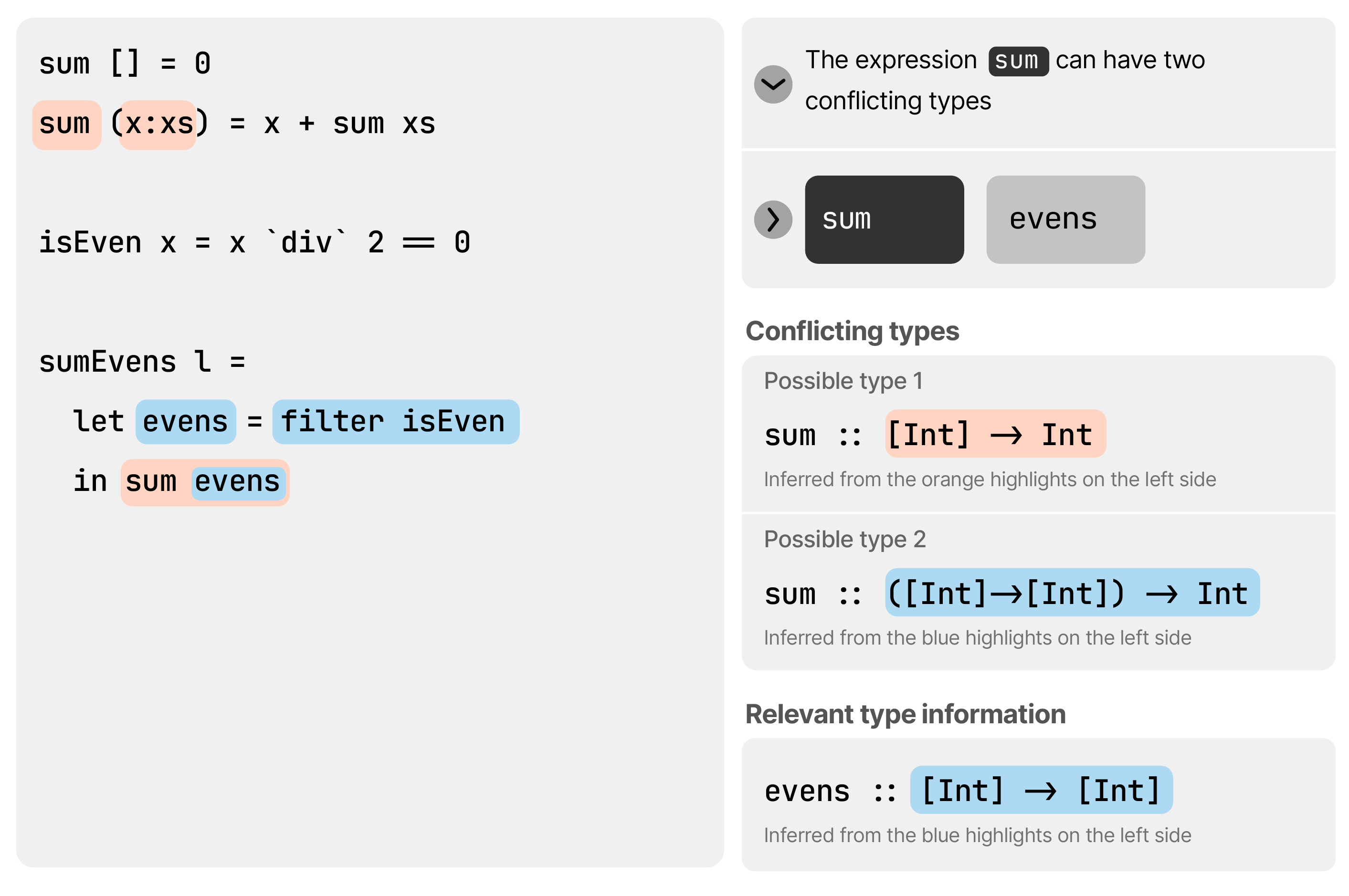}
        \caption{
            Maxine's code to calculate only the sum 
            of even numbers. \chameleon{} reports 
            an error with two candidate expressions.
        }
        \label{fig:balance-mode-1}
\end{figure}

Maxine writes additional code to add only even numbers in a list of integers, reusing the \texttt{sum} function she wrote earlier. After saving the file, \chameleon{} shows a type error in the expression \texttt{sum} (Fig.~\ref{fig:balance-mode-1}). However, this is not helpful because Maxine has just verified the implementation of \texttt{sum}. Switching to balanced mode, \chameleon{} shows two cards: \texttt{sum} and \texttt{evens}. 

\begin{figure}
   \centering
        \includegraphics[width=\linewidth,trim=0mm 8mm 0mm 0mm]{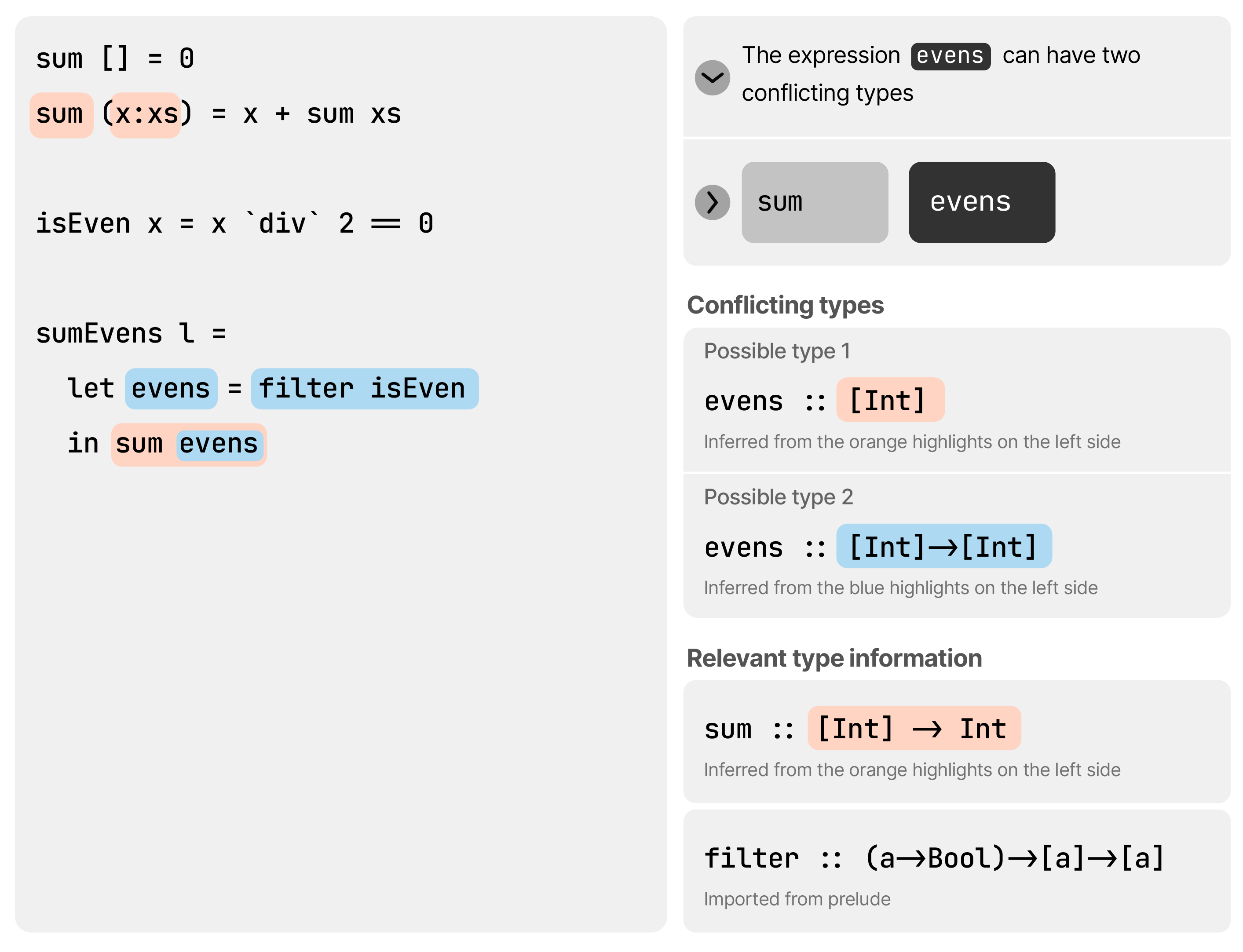}
        \caption{
            Clicking on the \texttt{evens} card (5) results in the changes in the
            conflicting types panel to show the possible types for \texttt{evens},
            and the changes highlight color to reflect the assumption that the
            definition of \texttt{evens} is the cause of the error.  
        }
        \label{fig:balance-mode-2}
\end{figure}

Maxine therefore clicks on the \texttt{evens} card and \chameleon{} reports two
possible types for the expression \texttt{[Int]} and \texttt{[Int] -> [Int]}
(Fig. \ref{fig:balance-mode-2}). Knowing the expression \texttt{evens} holds
a temporary list of even integers (hence it is of \texttt{[Int]} types), Maxine
concludes that the Possible type 2 is unintended. The locations with blue highlights must
contain the cause. It does not take long for Maxine to realize the list 
\texttt{l} is not supplied to the \texttt{filter} function.

\subsection{Advanced mode}  \label{sub:advanced}

\begin{figure}
        \centering
        \includegraphics[width=\linewidth,trim=0mm 8mm 0mm 0mm]{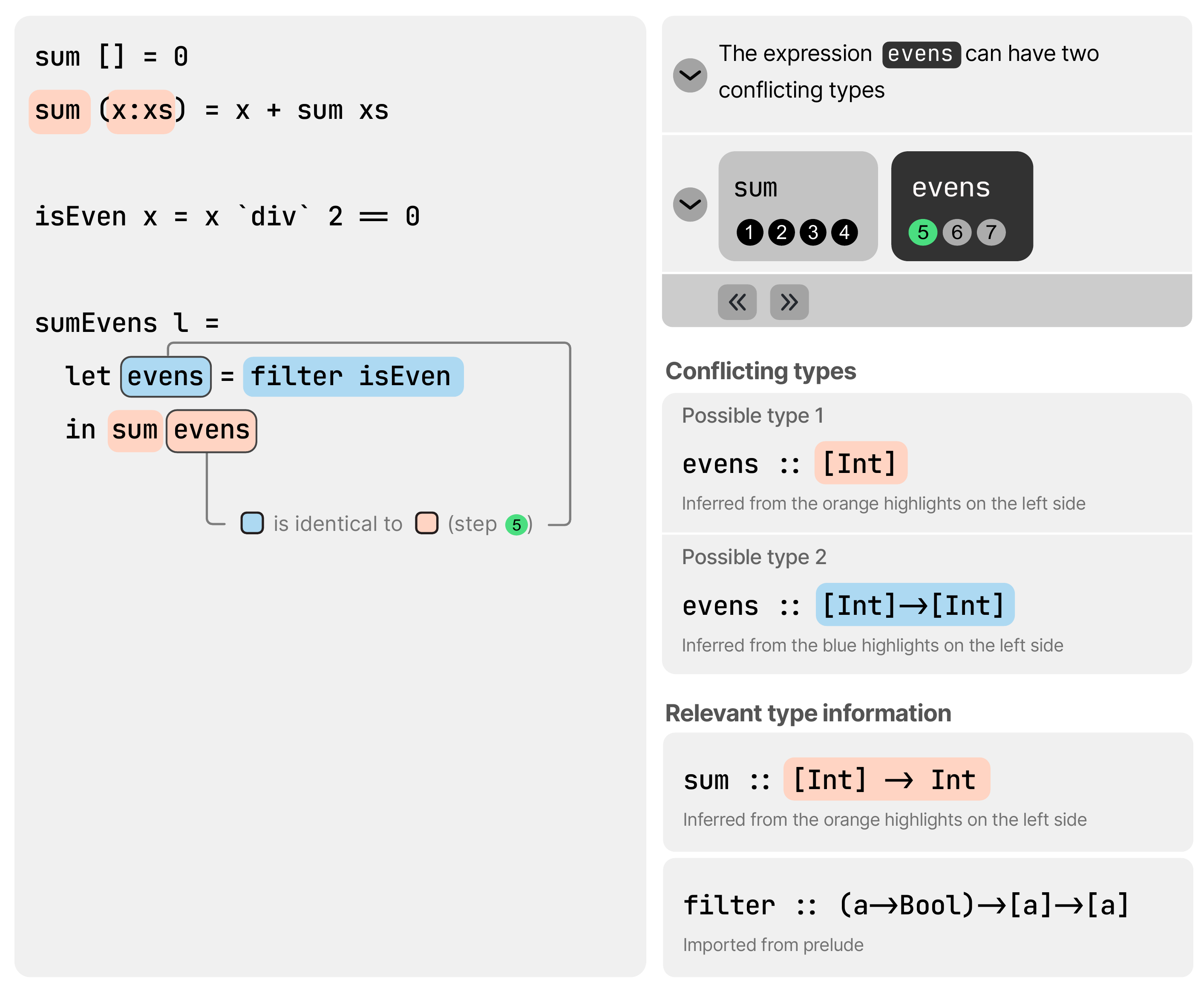}
        \caption{
            Maxine's code to calculate only the sum 
            of even numbers in advanced mode. 
            The current step is step 5, \chameleon{} 
            explains that the two appearances of expression 
            \texttt{evens} should have the same type.
        }
        \label{fig:advanced-mode-step5}
\end{figure}

\begin{figure}
        \centering
        \includegraphics[width=\linewidth,trim=0mm 8mm 0mm 0mm]{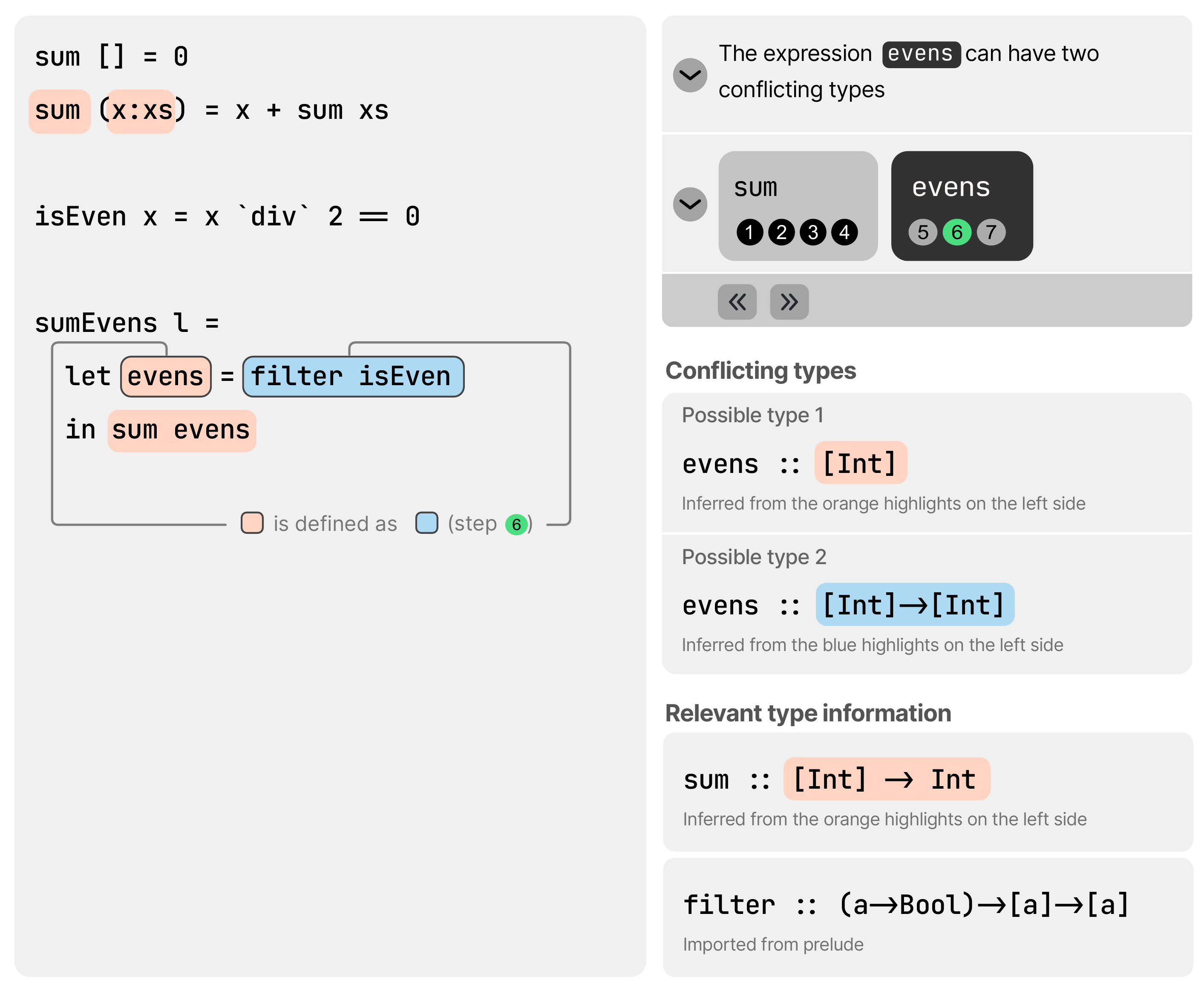}
        \caption{
            In step 6, \chameleon{} 
            explains that \texttt{evens} is defined as
            the expression \texttt{filter isEven}. The left-hand side
            and the right-hand side should have the same type.
        }
        \label{fig:advanced-mode-step6}
\end{figure}

\begin{figure}
        \centering
        \includegraphics[width=\linewidth,trim=0mm 8mm 0mm 0mm]{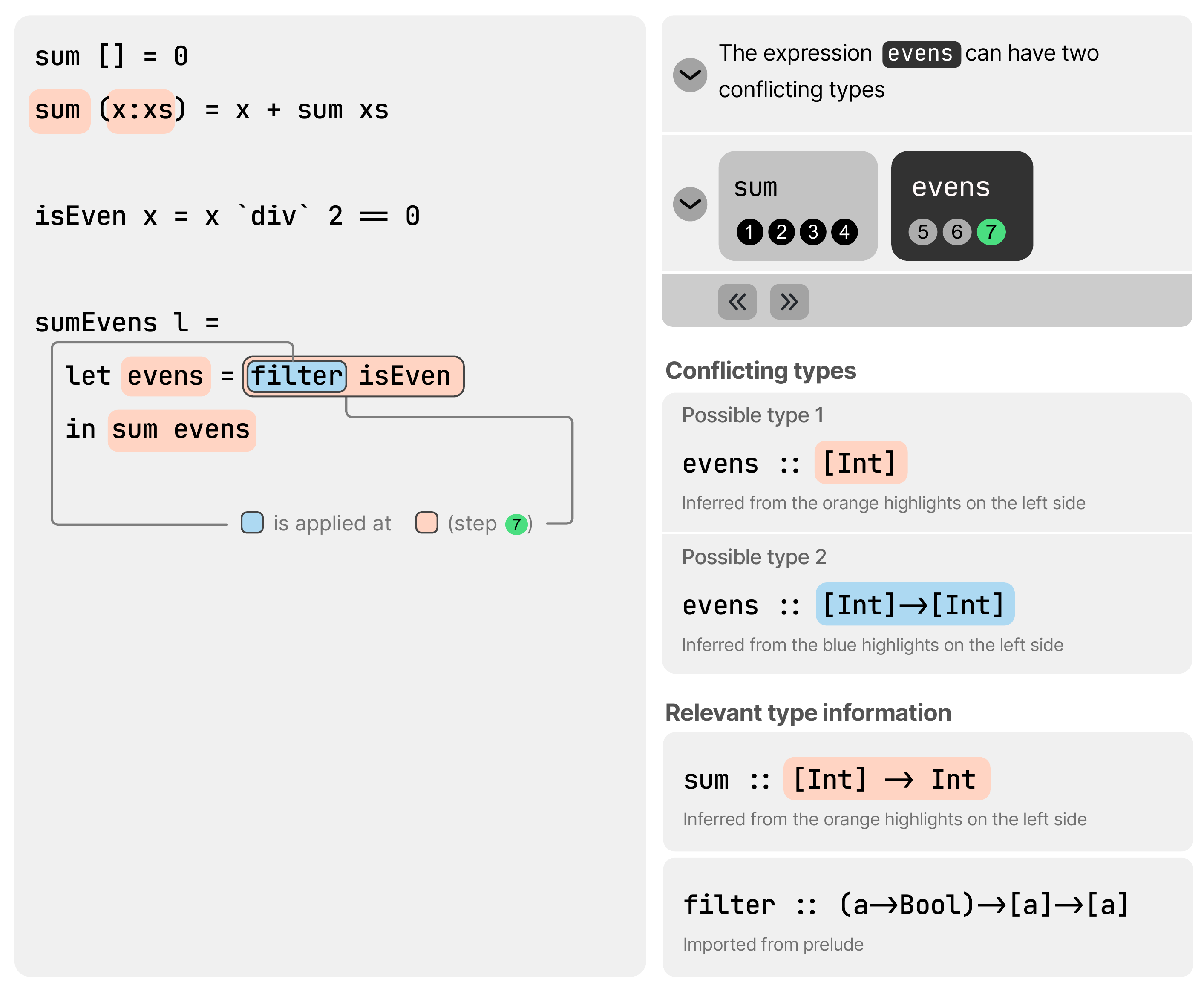}
        \caption{
            In step 7, \chameleon{} 
            explains that \texttt{filter} is applied to 
            the function \texttt{isEven}. Assisted by 
            the type of \texttt{filter} in the 
            Relevant Type Information panel on the bottom
            right, Maxine can find the type error that 
            \texttt{filter} expects two arguments but receives one.
        }
        \label{fig:advanced-mode-step7}
\end{figure}


To illustrate the deduction steps with the task shown in section \ref{sub:balanced},  first, Maxine clicks on step 5 (Fig. \ref{fig:advanced-mode-step5}) and verifies
that the two occurrences of \texttt{evens} are supposed to be identical, and the
second use means \texttt{evens} is a list of integers. Second, she
clicks on step 6 (Fig. \ref{fig:advanced-mode-step6}) and verifies that
\texttt{evens} should be the same type as the declaration on the right-hand
side.

Lastly, Maxine clicks on step 7 (Fig. \ref{fig:advanced-mode-step7}), and
it shows that the \texttt{filter} function is applied to one argument
\texttt{isEven}. By consulting the relevant type information, Maxine identifies
that \texttt{filter} is expecting two arguments while only one is provided.

\section{Evaluation}

\begin{figure}
    \centering
    \includegraphics[width=\columnwidth,trim=30mm 35mm 35mm 5mm]{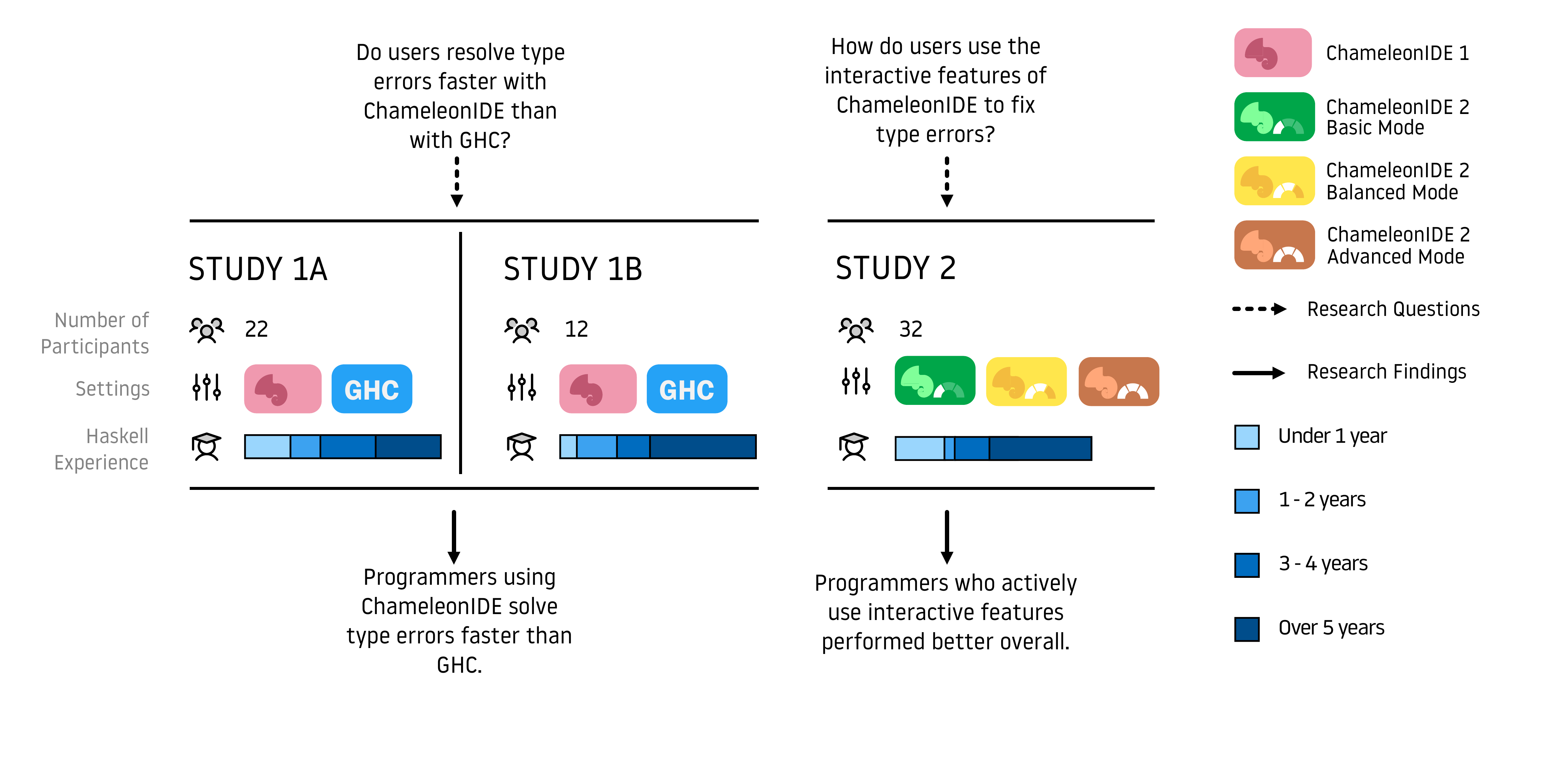}
    \caption{The timeline of \chameleon{}  evaluation.}
    \label{fig:timeline}
\end{figure}

We conducted three user studies, iteratively refining the \chameleon{} UI and evaluating several research questions as per Fig.~\ref{fig:timeline}. 

\subsection{Experiment Design}
\subsubsection*{\textbf{Recruitment}}

Participants were recruited via the Reddit \textit{r/haskell} and \textit{r/programminglanguages} communities. 
Participation is fully anonymized; detailed ethical implications of these experiments are reviewed and approved by the IRB of the authors' institution.

\subsubsection*{\textbf{Experiment setting}}
Experiments were conducted online and unsupervised. 
All user studies use a web-based debugging environment developed by the authors. 

\subsubsection*{\textbf{Training and group assignment}}
After consent, participants received interactive training on the tool interface and interactive features. Participants were also shown a cheat sheet summarizing the key functionality of the interface, and had access to the cheat sheet at all times during the study. Participants were given 4 trial runs (2 for each setting) before the data collection started. 
All the studies used a within-subject design to evaluate the effectiveness of different tools or feature sets while counterbalancing the difference in programming proficiency between participants. In each study, participants were required to complete a series of programming tasks (8 for studies 1a and 1b, 9 for study 2). At each task, a participant receives a single Haskell file that contains one or more type errors. They were then asked to correct the code with the help of the given tool.

\subsubsection*{\textbf{Data Collection}}
Time is measured from the start of each task to the first time the program is successfully type-checked and also passes all the functional tests. Participants are able to skip a task if they are stuck. 
After completing all tasks, participants are prompted to complete a debriefing survey. The survey questions include their Haskell experience and feedback on the tools.

We used a browser session recording tool~\cite{openreplay_openreplay_2022} to record the study sessions. This allows us to identify usability issues in the study and to recognize general patterns. 

\subsection{\chameleon{} Human Studies}

\subsubsection{\textbf{\chameleon{} 1}}  
An earlier version of the UI than that depicted in Figs.~(2-13), it featured the type inference engine that recovers most concrete types after type errors occur and a minimal set of debugging features. Key features in \chameleon{} 1 include showing two (or more) alternative types, showing all possible error locations, dividing possible error locations into groups based on alternative types, and concrete type restoration. In short, \chameleon{} 1 is equivalent to \chameleon{} 2 set to basic mode. 


Two  studies (1a \& 1b) were conducted to compare the effectiveness of solving type errors using \chameleon{} 1 and GHC compiler error messages. We choose GHC compiler error messages as the baseline because it is the canonical tool for working with type errors in Haskell.

Eight tasks were given in both studies. In study 1a, the tasks were taken from the exercises of the Haskell programming class in the authors' institute. In the second study, the tasks are sourced from the top 20 Haskell topics on GitHub~\cite{github_github_2022}. The authors then manually added type errors into the program. In both studies, the type errors include simple mismatch, confusing syntax, missing instance, precedence and fixation, infinite types, and confusing list versus element. These categories follow the common type errors in Tirronen's study \cite{tirronen_understanding_2015}. 

Studies (1a \& 1b) address the research question:

\noindent\textbf{RQ1.} \textit{Do programmers solve type errors faster with \chameleon{} than GHC compiler error messages?}




\begin{figure}
    \centering
    \includegraphics[width=0.8\linewidth,trim=15mm 12mm 15mm 35mm,clip]{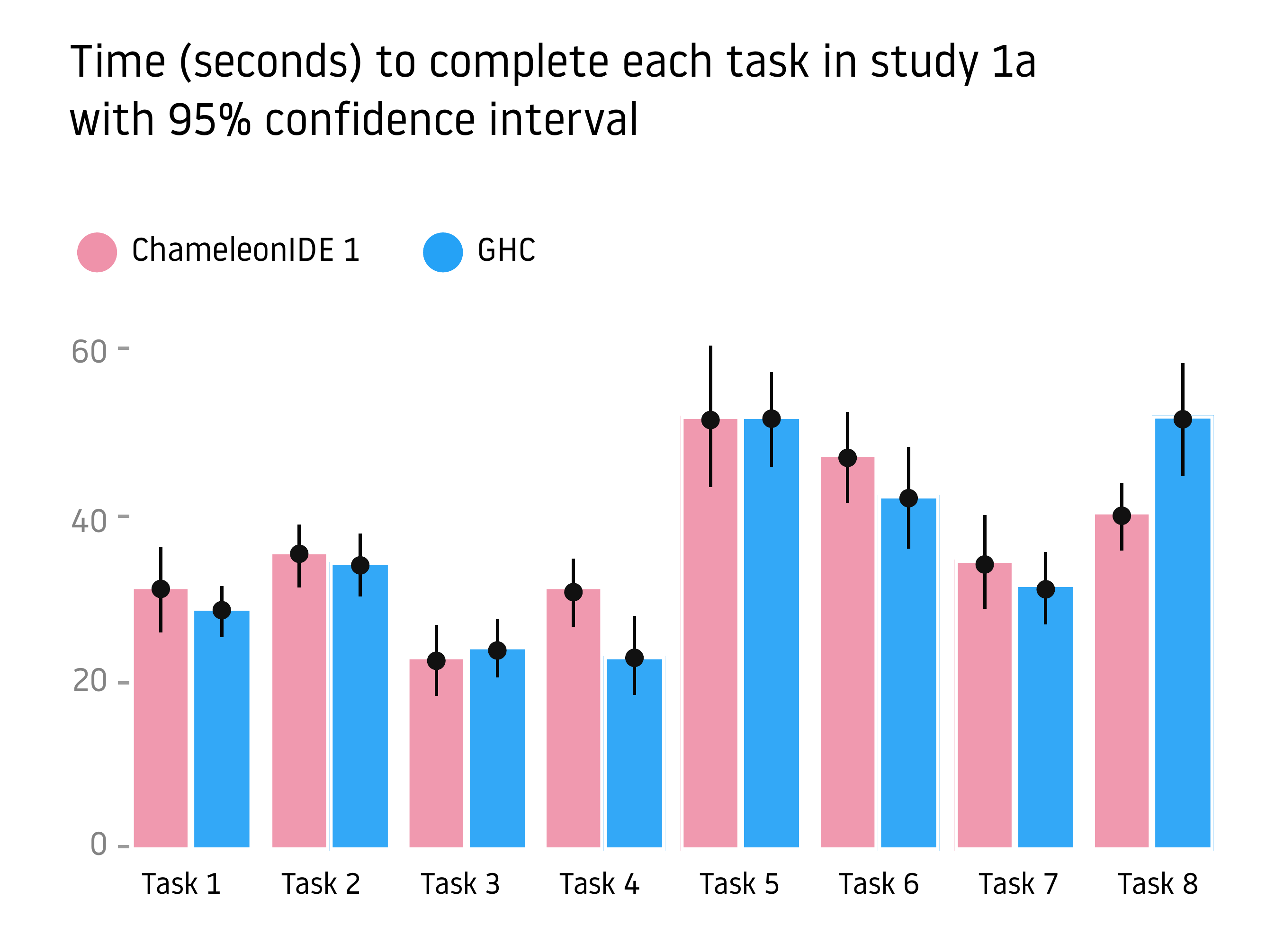}
    \caption{Study 1a task completion time (secs.) with 95\% confidence interval.}
    \label{fig:analysis-1a}
\end{figure}

\begin{figure}
    \centering
    \includegraphics[width=0.8\linewidth,trim=12mm 15mm 12mm 35mm,clip]{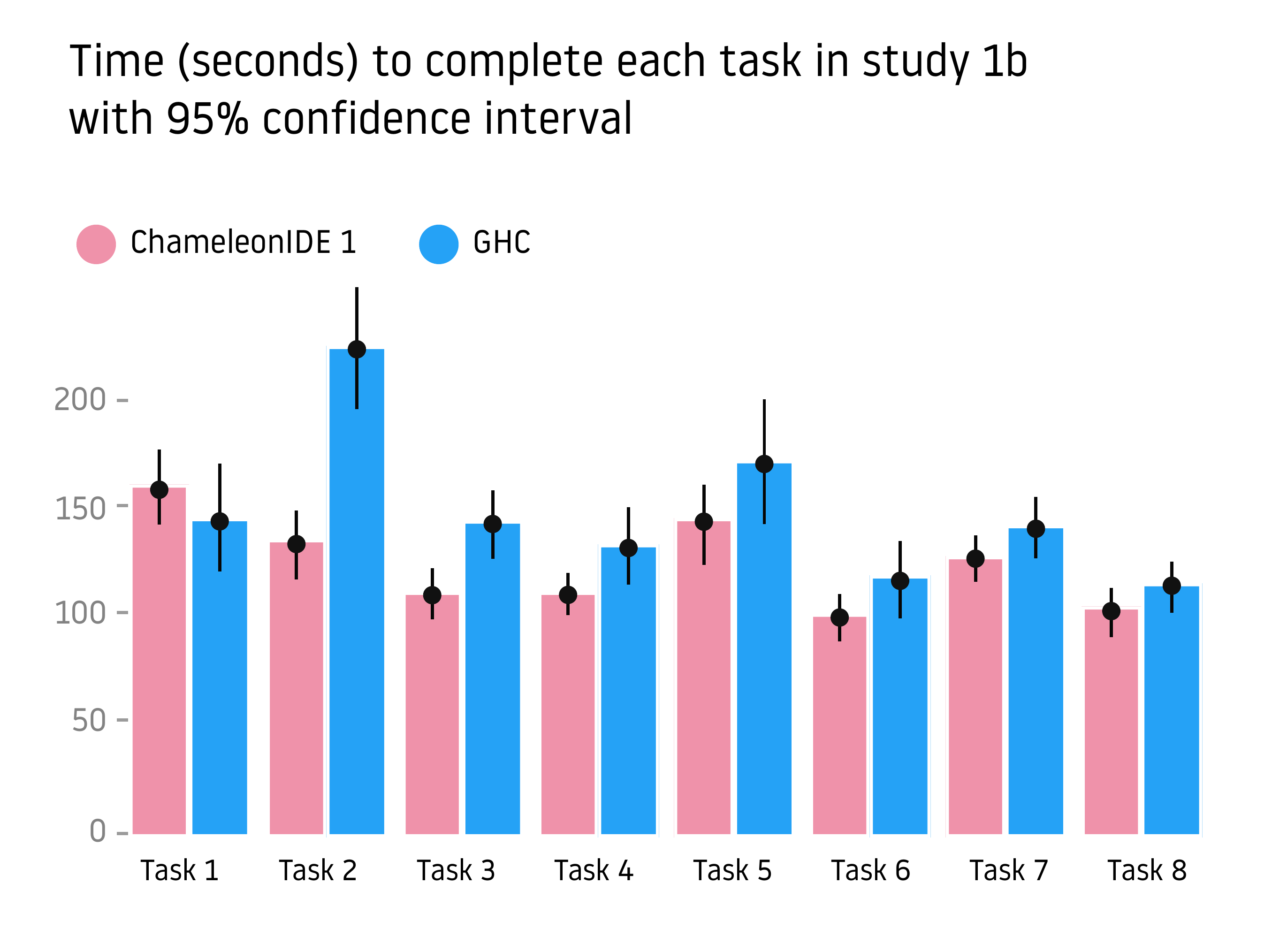}
    \caption{Study 1b task completion time (secs.) with 95\% confidence interval.}
    \label{fig:analysis-1b}
\end{figure}

\subsubsection*{\textbf {Results}}

The data collected during study 1a, Fig.~\ref{fig:analysis-1a} does not show significant differences across Tasks 1-7. In hindsight, these tasks were trivial challenges for most users, and the individual differences among participants are generally more significant than the differences between treatments. However, one interesting observation is task 8, where the \chameleon{} group outperformed the GHC group. We attribute this significant difference to the difficulty of Task 8. The source file is longer and involves more language features (abstract data types and high-level functions). GHC struggles to produce a relevant error message for this type of error. 
From this result, we hypothesized that we might observe a more significant difference using tasks with lengthier and more realistic source code. 
This hypothesis is also supported by the most common feedback claiming that the tasks were too trivial to invite meaningful evaluation. One participant said, ``Looks nicer than GHC, but without trying it on something more complicated, I cannot conclude whether it would help me in practice." 

Therefore, in study 1b we introduced more difficult challenges and indeed observed that the \chameleon{} group was faster than the GHC group in almost all tasks (figure \ref{fig:analysis-1b}), barring task 1. A two-sample paired t-test was performed to compare the completion time between \chameleon{} and GHC groups. There was a significant difference between the two groups: $t(23) = -3.86, p = 0007$. For task 1, it is suspected that some participants spent more time exploring the interface of \chameleon{} due to its unfamiliarity. For all other tasks, from the video recordings, we saw many \chameleon{} users confidently skip reading unrelated chunks of code, while GHC users generally read through the whole program. In harder problems and messier code, we notice programmers start to report the benefits of \chameleon{}. ``It's most useful feature that I noticed was that it points out the locations of both conflicting uses; GHC often makes it difficult to figure out how it's coming to a conclusion about a type." reported one participant. ``I think \chameleon{}  does a much better job than GHC's error messages. I like that it shows the sources for the type judgments. This makes it quite easy to figure out how to rectify errors." reported another participant.

\subsubsection{\textbf{\chameleon{} 2}}  \label{sub:us4}
Based on observations of Study 1 we introduced several new features to \chameleon{}, eventually resulting in the UI depicted in Figs.~(2-13). Interactive features were available in this iteration, such as deduction steps, candidate expressions, and mode switching. A few other user interfaces \cite{anonymous_interactive_2021} were designed and prototyped between the development of \chameleon{} 1 and \chameleon{} 2. Study 2 addresses the research question: 

\noindent\textbf{RQ2:} \textit{How do programmers use the interactive features in \chameleon{} 2?}. 

More specifically:
\begin{itemize}
    \item \textbf{RQ2.1} How do programmers use the advanced features provided by \chameleon{} 2?
    \item \textbf{RQ2.2} Do programmers prefer switching modes during debugging type errors?
    \item  \textbf{RQ2.3} What are programmers' preferences among the three modes provided by \chameleon{} 2?

\end{itemize}

During each run, the initial mode of each task alternated through the three different modes and repeated three cycles in nine tasks. The order of the three modes in each cycle is counterbalanced among all participants. However, participants can switch to other modes at any time.


\subsubsection*{\textbf{Results}} 

Study 2 is more exploratory in methodology than Study 1. We encouraged programmers  to discover their way of using the tool. In post hoc analysis of the collected log data, we were able to extrapolate some interesting patterns of how the tool was used.

\textbf{RQ2.1}. The most striking feature of the data is that users tend to vary wildly in their use of the tool. Some users used the features extensively, while others completed the tasks without actively exploring the given information. Based on this discrepancy, we divided the users into three groups in table~\ref{tab:interaction-level}.
\begin{table}
    \centering
\begin{scriptsize}
\begin{small}
\noindent\begin{tabularx}{\linewidth}{ 
  | >{\hsize=.26\hsize}X 
  | >{\hsize=.74\hsize \raggedright\arraybackslash}X  | }
    \hline
        Interaction level & Description \\ \hline
        \textit{Minimal}  & Users completed the tasks by making changes in source code, type checking, and reading error messages. \\ \hline
        \textit{Low}  & Users only actively used universal features in all modes, for example, hovering on "Possible type 1" and "Possible type 2" to narrow down error space. \\ \hline
        \textit{High}  & Users did everything from the low interaction group but used features specific to the Balanced mode and the Advanced mode, such as activating steps and expression cards. \\ \hline
\end{tabularx}
\end{small}
\end{scriptsize}
    \caption{Levels of programmer interaction and their description}
    \label {tab:interaction-level}
\end{table}

\begin{figure}
    \centering
    \includegraphics[width=\linewidth,trim=0mm 15mm 0mm 50mm,clip]{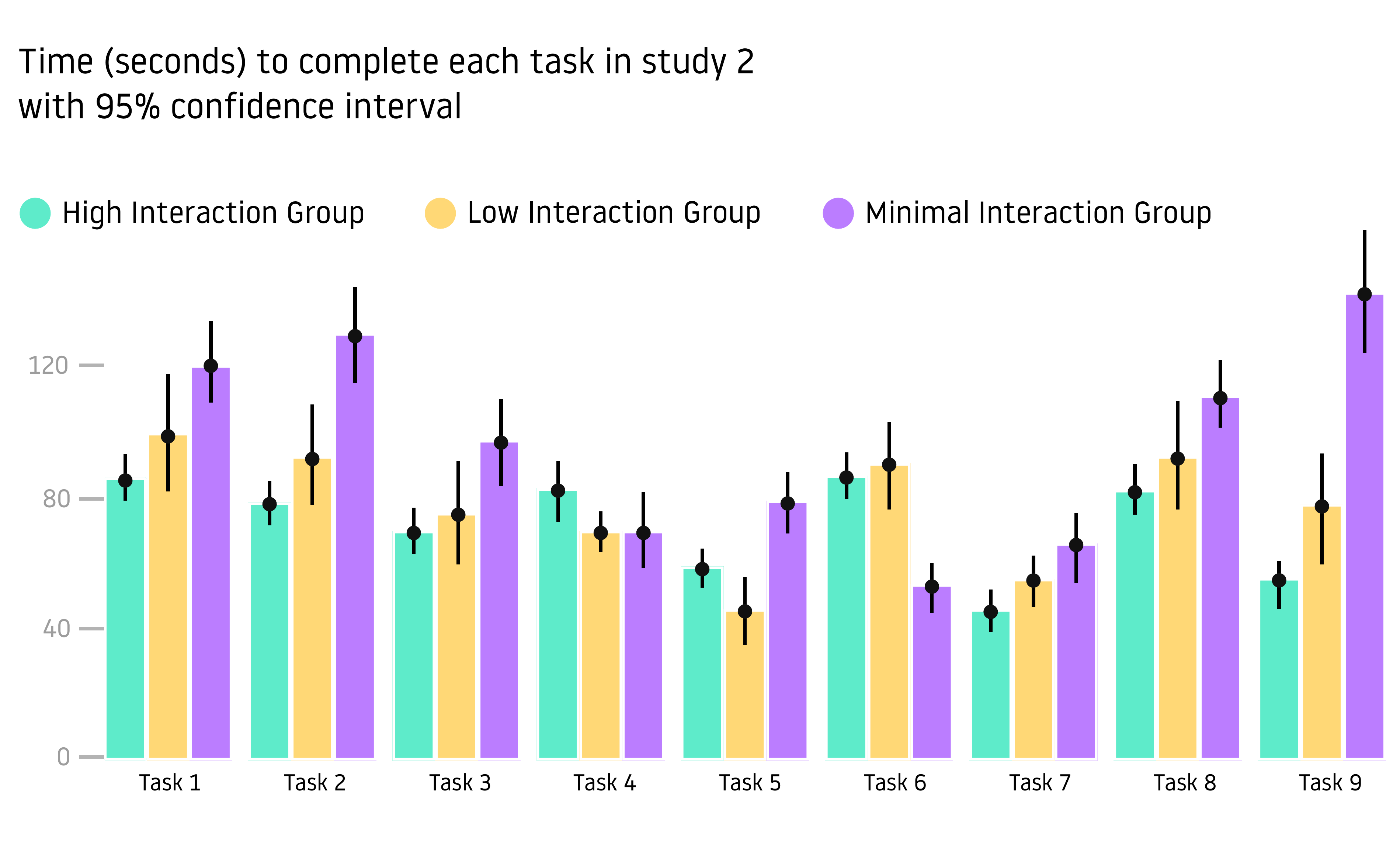}
    \caption{Study 2 task completion time (secs.) with 95\% confidence intervals.}
    \label{fig:r4-analysis}
\end{figure}

As shown in  Fig.~\ref{fig:r4-analysis}, the time to complete each task roughly relates to the interaction level of participants. Participants with higher interaction levels generally performed better, and the lowest interaction level was worse. Tukey’s HSD Test for multiple comparisons found that the completion time was significantly different between the minimal interaction group and the high interaction group ($p \le 0.001$, 95\% C.I. = [18.26, 31.41]), and between the minimal interaction group and the low interaction group ($p \le 0.001$, 95\% C.I. = [11.96, 26.67]). The results from three tasks stand out from the general trend: in Tasks 4 and 6, higher interaction users performed worse, and in task 9, the general trend is exaggerated. As with Study 1a and 1b, this difference is likely related to task difficulty. Tasks 4 and 6 are shorter than other tasks. The ideal fixes for these two tasks are placed relatively early in the source code (both in the first two lines of the source code). Users simply reading top to bottom could quickly identify the error without needing to skip unrelated sections of code using the information provided by \chameleon{}. This reduced the apparent benefit of \chameleon{} in these tasks. On the other hand, task 9 is the lengthiest task of all. It also involves deeply nested type definitions that are harder to follow in mind.

\begin{figure}
    \centering
    \includegraphics[width=0.7\linewidth,trim=0mm 5mm 0mm 5mm,clip]{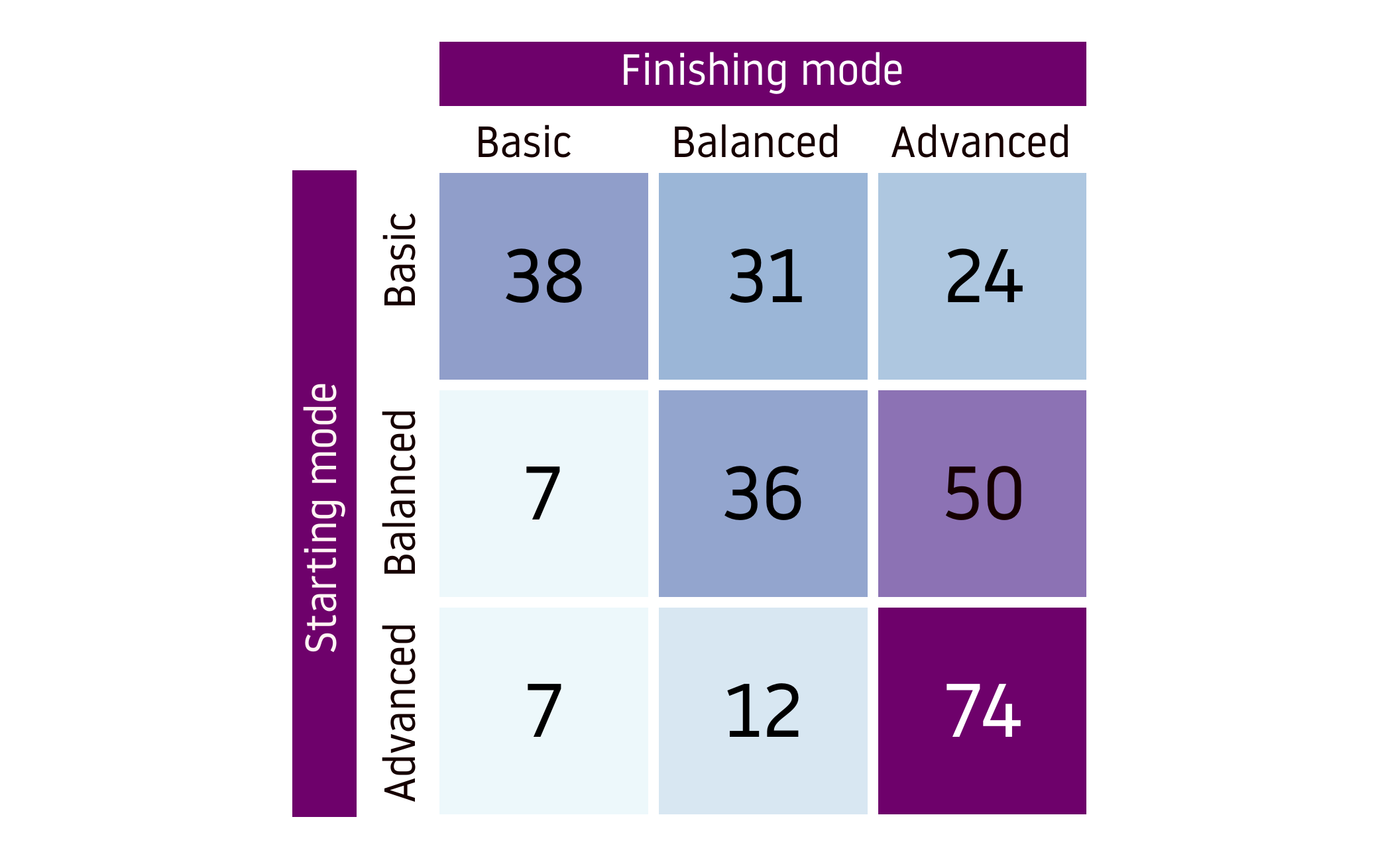}
    \caption{Study 2 mode switches by starting mode.  Users overwhelmingly switched to the more sophisticated interface mode.
    }
    \label{fig:r4-mode-switching}
\end{figure}

Another observation is when using the mode switching feature of \chameleon{}, we show this by presenting the starting mode and finishing mode of each task and each participant in a correlation matrix (Fig. \ref{fig:r4-mode-switching}). This observation suggests two characteristics of using multi-mode debugging tools. First, to answer \textbf{RQ2.2} programmers are roughly splitted in this matter: 53\% changing modes vs. 47\% staying in the same mode. Second, to answer \textbf{RQ2.3} when changing modes, programmers generally switch to the more informative modes instead of the more concise ones.

\subsection{Limitations}

One threat to the validity of the evaluation is the number of participants. Although for each study we received hundreds of online participants, the studies suffered from a high abandonment rate (especially study 1b). This was expected: the programming challenges are difficult, and our volunteer participants are unremunerated. 
Because we recruited participants online and anonymized all the participants, it is possible for participants of a previous study to enter a later one. This creates variation in familiarity. We offset this by using new code challenges in every study and conducting trial runs before data collection to bring new participants up to speed.
Conducting studies remotely and unsupervised left us no means to intervene when users encounter usability issues. To mitigate this, we conducted cognitive walkthroughs and sandbox pilots before running each study.

Future evaluation would benefit from using more realistic tasks. The tasks in our human studies do not get as complex as professional Haskell programmers may face in a typical production codebase. It would be interesting to see how \chameleon{} is used against type errors that span multiple files and packages and include more confusing abstractions, like Monads, Monad transformers, and Lenses.

\section{Discussion}


This paper presents the interactive type debugging tool \chameleon{} and charts the evolution of its design across several iterations in response to user evaluation and feedback, as well as examines   the effectiveness of the general approach compared to traditional static type error messages. We found that programmers using \chameleon{} are able to debug errors faster than using traditional text-based error messages. This effect is shown more clearly when the task is not trivial. We found that programmers who actively use \chameleon{}'s interactive features are more efficient in fixing type errors than passively reading the type error output. In this section, we will discuss a few interpretations of the results.

\subsection{Effect on Reading Source Code}
From the results of Study 1a, we observed that the choice of debugging tool had little effect on how fast programmers solve simple type errors. Conversely, when facing more realistic problems (longer source code, error locations more scattered) in study 1b, programmers are more effective using \chameleon{}. One explanation is that \chameleon{} reduces the amount of reading time by taking programmers more directly to the problem. Earlier studies \cite{jbara_how_2015, peitek_what_2020} showed that reading source code is generally the initial step of solving programming problems and is done in several passes. Although traditional compiler error message tools initially show fewer locations, these may be incomplete, meaning that programmers have to expand the reading span without clear guidance. In contrast, \chameleon{} shows more error locations initially. However, the completeness of error locations assures programmers which part of the source code can be safely skipped.

\subsection{Forming Debugging Plans}
From the results of Study 2, we found that programmers who use the interactive tool fix type errors faster than the ones who passively read the error output. This effect is stronger in harder tasks. We speculate that one factor of this result is that  \chameleon{} helps to develop debugging plans. We observed that when working with \chameleon{}, programmers form different debugging plans to attack the problem. Among the \textit{high} interactivity participants in user study 2, some programmers cycle through deduction steps as a guide to reading source code; some navigate to both ends of the deduction chain where types are normally grounded and concrete. In contrast, \textit{minimal} interactive participants generally form similar plans, including carefully reading the program text and manually annotating expressions based on their understanding of the program.

\subsection{Externalize Intermediate Typing Information}
We speculate another factor of the effectiveness of \chameleon{} interactive debugging tools is that they help programmers effectively chunk intermediate information. With the program shown in Listing~\ref{listing:2}, \chameleon{} offers two candidate expressions: \texttt{f} can be typed as \texttt{Int -> Bool} or \texttt{Char -> Bool}; \texttt{z} can be typed as \texttt{Int} or \texttt{Char}. Although  these two statements are equivalent in theory, programmers are often required to compute the latter from the former or vice versa. And this computation may carry out multiple layers. Programmers have to remember all the intermediate types and their reasoning throughout such mental gymnastics. Assisted by candidate expression cards and deduction steps, this intermediate information is externalized on screen and can be retrieved anytime. A recent study on working memory \cite{crichton_role_2021} suggested this approach may provide a positive effect in helping programmers manage cognitive load and free up working-memory space for high-level thinking.




\begin{listing}
\begin{Verbatim}[commandchars=\\\{\}]
    \PYG{n+nf}{f} \PYG{n}{z}
        \PYG{o}{|} \PYG{n}{z} \PYG{o}{==} \PYG{l+m+mi}{3} \PYG{o+ow}{=} \PYG{k+kt}{False}
        \PYG{o}{|} \PYG{n}{z} \PYG{o}{==} \PYG{l+s+sc}{\PYGZsq{}4\PYGZsq{}} \PYG{o+ow}{=} \PYG{k+kt}{True}
\end{Verbatim}
\vspace*{-2mm}
\caption{\chameleon{} reports an error in the expressions \texttt{f} and \texttt{z}}
\label{listing:2}
\end{listing}

\section{Related Work}



\subsection{Finding all type error locations}
Many have studied the approach of finding all locations that contribute to a type error~\cite{stuckey_interactive_2003, haack_type_2004, pavlinovic_practical_2015, schilling_constraint-free_2012}. Type error slicing~\cite{haack_type_2004} is a technique that finds locations that are complete and minimal for the type error. Internally labeled constraints and Minimal Unsatisfiable Subset (MUS) generation are used to generate these slices. The language supported in Haack's work was a subset of Standard ML. The original Chameleon~\cite{stuckey_interactive_2003} used  Constraint handling rules (CHR) to support the computing of type error slices in Haskell. Chameleon also supported advanced type-level features (type classes and functionally dependent types). The project also introduced the ability to query type information through a command line interface. Although Chameleon was firmly grounded in results from type theory, its designs were never evaluated with user studies. While finding all error locations is useful in comprehending type errors, it is only 1 of the 7 properties listed in the proposed manifesto of good type error reporting~\cite{yang_improved_2000}.
To the best of our knowledge, ours is the first user-centered evaluation of an interactive type debugging system involving type-error slicing.

\subsection{Producing high-quality error explanation}
One weakness of compiler error messages, in general, is that they fail to explain the error in human language. As put in~\cite{barik_developers_2017}, ``Error messages appear to take the form of natural language, yet are as difficult to read as source code."  A well-studied approach to producing better error explanations is through ECEM (Enhanced compiler error message). Through a series of mixed-method studies, Prather showed~\cite{prather_novices_2017} that ECEM has a positive result in understanding compiler errors. Decaf~\cite{becker_effective_2016} is a tool that can rephrase Java compiler error messages into an enhanced version. In a study of over 200 CS1 students, Decaf was shown to reduce overall errors in their coding practices. Berik proposed a framework~\cite{barik_how_2018} for constructing compiler error messages based on argumentation theory, and showed that error messages following a simple argumentation layout or an extended argumentation layout are more human-friendly.  These works show the significance of improving the language in the compiler error messages. Most principles and suggestions are followed in \chameleon{} in constructing error statements. However, these earlier studies were not targeting type errors alone but general compiler errors (some even include runtime errors). The nuances of type errors, such as alternative typing, were not considered. Moreover, these explanation systems were designed specifically for novice users.


\subsection{Interactive Debugging}

Modern programming tools can offer alternative methods of code authoring, display real-time feedback and reveal complex programming contexts through visualizations. Many tools aim to improve the debugging experience using such capabilities. We list two. Hazel Tutor \cite{potter_hazel_2020} is an interactive type-driven environment for the OCaml language. It can automatically fill type holes by suggesting template expressions (called ``strategies'' by the authors) through a popup window. It also provides a cursor-based type inspector that allows programmers to query the types of different parts of the program. Whyline~\cite{ko_finding_2009} is a Java debugging system that allows a user to ask questions like ``why does variable X have value Y." It also allows users to interactively ask follow-up questions to gain further knowledge of the nature of an error. These debugging tools are important motivations for developing \chameleon{}. However, they focus on different aspects of the debugging process. Java Whyline mainly tackles the problem of unintended runtime behavior, while Hazel Tutor specializes in development assistance supported by type holes.


\section{Conclusion}

We present \chameleon{}, a type debugging tool for the Haskell programming language. Its constraint-based type inference engine provides unbiased and comprehensive error location reporting. 
Our studies evaluated the tool's design with programmers. We found that, particularly for more complex tasks, \chameleon{} helped programmers to fix type errors more quickly than traditional text-based error messages. Further, programmers actively using \chameleon{} interactive features are shown to fix type errors faster than simply reading the type error output.
\chameleon{} currently works with the Haskell language, but in the future, we plan to extend the type-checking system to work with other strongly typed languages, such as Rust or TypeScript.

\subsection*{Acknowledgments} 

The work of Peter Stuckey was partially supported by the OPTIMA ARC ITTC, Project ID IC200100009.

\goodbreak\noindent

\bibliographystyle{IEEEtran}
\bibliography{ICPC}

\end{document}